\documentclass[%
 aip,
 amsmath,amssymb,
 reprint,%
]{revtex4-1}

\usepackage{graphicx}
\usepackage{dcolumn}
\usepackage{bm}

\usepackage{hyperref}
\usepackage[utf8]{inputenc}
\usepackage[T1]{fontenc}
\usepackage{mathptmx}
\usepackage{etoolbox}

\makeatletter
\def\@email#1#2{%
 \endgroup
 \patchcmd{\titleblock@produce}
  {\frontmatter@RRAPformat}
  {\frontmatter@RRAPformat{\produce@RRAP{*#1\href{mailto:#2}{#2}}}\frontmatter@RRAPformat}
  {}{}
}%
\makeatother
\begin{document}

\preprint{AIP/123-QED}

\title[]{Quantum Error Mitigation with Attention Graph Transformers for Burgers Equation Solvers on NISQ Hardware}
\author{Seyed Mohamad Ali Tousi}

 \affiliation{ 
Vision-Guided and Intelligent Robotics Lab (ViGIR) - EECS Department - University of Missouri - Columbia 
}%

\author{Adib Bazgir}%
 
\affiliation{ 
Department of Mechanical and Aerospace
Engineering - 
University of Missouri - Columbia 
}%

\author{Yuwen Zhang}
\homepage{Corresponding Author: zhangyu@missouri.edu}
\affiliation{ 
Department of Mechanical and Aerospace
Engineering - 
University of Missouri - Columbia 
}%

\author{G. N. DeSouza}
 \homepage{Corresponding Author: DeSouzaG@missouri.edu}
\affiliation{ 
Vision-Guided and Intelligent Robotics Lab (ViGIR) - EECS Department - University of Missouri - Columbia 
}%

\date{\today}

\begin{abstract}
We develop a hybrid quantum–classical framework, augmented by learned error mitigation, to solve the viscous Burgers equations on noisy intermediate-scale quantum (NISQ) hardware. The nonlinear Burgers equation is mapped via the Cole–Hopf transform to a diffusion equation, discretized on uniform grids and encoded in a quantum state whose time evolution is approximated by Trotterized nearest-neighbor $R_{XX}$ circuits implemented in Qiskit. Quantum simulations are performed on noisy Aer simulation backends and IBM superconducting devices and are benchmarked against high-accuracy classical solutions obtained from a Krylov-based solver applied to the corresponding discretized Hamiltonian. From measured quantum amplitudes, we reconstruct the velocity field $u(x,t)$ and evaluate physical and numerical diagnostics including $L^{2}$ error, shock position, and dissipation rate, with and without zero-noise extrapolation (ZNE). To enable data-driven error mitigation, we generate a large parametric dataset by sweeping viscosity, time step, grid size, and boundary conditions, yielding matched tuples of noisy, ZNE-corrected, hardware, and classical solutions together with detailed circuit metadata. On top of this dataset, we train an attention graph neural network that ingests circuit and light-cone features, global circuit parameters, and noisy quantum outputs and learns to predict an error-mitigated solution that closely approximates the classical reference. Across diverse parameter settings, the learned model systematically reduces the discrepancy between quantum and classical solutions beyond what is achieved by ZNE alone. We outline how this pipeline extends to higher-dimensional Burgers systems and more general quantum PDE solvers, and argue that learned error mitigation provides a promising complement to physics-based noise correction on NISQ devices.
\end{abstract}

\maketitle

\section{\label{sec:level1}Introduction}

The one-dimensional viscous Burgers equation
\begin{equation}
\frac{\partial u}{\partial t}
+ u \frac{\partial u}{\partial x}
= \nu \frac{\partial^{2} u}{\partial x^{2}}, 
\qquad \nu > 0,
\end{equation}
is a canonical model for nonlinear advection--diffusion and has long served as a testbed for numerical methods, turbulence modelling, and flow control. Historically, Bateman and Burgers introduced convective--diffusive models of this form as simplified analogues of the Navier--Stokes equations that retain key features such as nonlinearity, shock formation, and viscous regularization. Comprehensive historical and methodological reviews emphasize both the mathematical structure and physical relevance of Burgers-type equations across gas dynamics, sedimentation, traffic flow, and interface growth, as well as the challenges of designing accurate, stable, and convergent schemes at low viscosity and in the presence of steep gradients \cite{dhawan2012contemporary,bonkile2018systematic}. Classical computational fluid dynamics (CFD) for Burgers and related PDEs is now highly mature, with finite-difference, finite-volume, spectral, and stabilized Galerkin schemes routinely used in engineering and physics \cite{basu2025solving}.

Despite this maturity, high-fidelity simulations at large Reynolds numbers remain computationally demanding, as resolving sharp gradients and small dissipative scales requires large grids and small time steps. Even with mesh adaptivity and reduced-order modelling, the cost of fully resolving nonlinear transport and turbulence scales unfavourably with problem size \cite{tennie2025quantum}. At the same time, the effective slowdown of Moore’s law and the plateauing of conventional high-performance computing have motivated exploration of alternative paradigms, most prominently quantum computing, for PDEs, nonlinear dynamics, and turbulence. Recent perspectives on quantum computing for nonlinear differential equations and turbulence explicitly argue that new hardware and algorithmic synergies will be needed to tackle multiscale flows beyond the reach of classical resources, and they highlight Burgers-like equations as prototypical testbeds in this context \cite{amaral2025quantum,gonzalez2025quantum}.

Within quantum computing, Burgers’ equation has naturally emerged as a focal point for early quantum PDE algorithms. Yepez et al. developed quantum lattice-gas and quantum lattice--Boltzmann models in which occupation amplitudes evolve under unitary collision--streaming rules whose hydrodynamic limits recover viscous Burgers and, more generally, Navier--Stokes-type equations \cite{yepez2002efficient,yepez2006open}. 
In these schemes, local collision operators map directly to few-qubit gates and streaming corresponds to qubit permutations, giving a transparent mapping between quantum circuits and hydrodynamic observables. Complementary approaches build on quantum PDE algorithms in the spirit of Gaitan et al. to solve viscous and inviscid Burgers equations, demonstrating numerically that quantum solutions converge to shock profiles and travelling waves with accuracy comparable to reference CFD schemes while preserving the theoretical complexity advantages of the underlying quantum algorithm. Related studies extend these methods to advection--diffusion equations and Burgers-type flows under various initial and boundary conditions, systematically exploring the impact of central differencing, upwinding, and Lax--Wendroff schemes on numerical dissipation and dispersion, and recovering rarefaction waves and expansion shocks in the quantum setting \cite{oz2022solving,lubasch2020variational}. Steijl further showed that carefully designed quantum circuits can emulate a lattice--Boltzmann discretization of Burgers, making explicit the mapping between collision operators, unitary gates, and hydrodynamic observables \cite{steijl2022quantum}. Together, these works establish Burgers as a structured benchmark for quantum fluid solvers ranging from quantum lattice models to more general quantum PDE constructions. 

Beyond application-specific circuit designs, a complementary line of work has sought general quantum algorithms for nonlinear PDEs that specialize to Burgers’ equation. Liu et al. propose an efficient Carleman-linearization-based quantum algorithm for dissipative nonlinear ODEs and PDEs, explicitly using Burgers’ equation as a testbed to investigate truncation error, dissipation conditions, and nonlinearity \cite{liu2021efficient}. Building on this, Wu et al. show that Carleman linearization can converge under resonance conditions without strong dissipation assumptions, expanding the class of nonlinear PDEs including Burgers- and KdV-type equations amenable to quantum advantage \cite{wu2025quantum}. Setty develops block-encoding and quantum singular-value-transformation (QSVT) techniques for Carleman-linearized transport PDEs, with Burgers dynamics serving as a key benchmark for analysing circuit depth and minimum-singular-value scaling \cite{setty2025quantum}. Demirdjian et al. further decompose the Carleman-linearized one-dimensional Burgers equation into a polylogarithmic number of block-encodable terms, enabling variational quantum linear solvers (VQLS) to access truncated Burgers systems with controlled two-qubit gate depth \cite{demirdjian2025efficient}. At a more formal level, Gonzalez-Conde et al. study quantum Carleman linearization for nonlinear fluid dynamics and show how physical flow parameters and grid resolution control the efficiency of the linearization and the resulting quantum algorithm \cite{gonzalez2025quantum}. These developments clarify both the theoretical scaling of quantum PDE solvers and the role of dissipation and resonance in ensuring controllable truncation errors for nonlinear transport phenomena. 

In parallel, variational and hybrid quantum--classical methods have been proposed to make differential-equation solvers more compatible with noisy intermediate-scale quantum (NISQ) hardware. Schilling and Sturm formulate variational quantum algorithms for differential equations on noisy processors, emphasizing shallow ansätze, time-discretized objective functions, and robust cost constructions \cite{schillo2025variational}. 
Over et al. extend this framework to generalized linear and nonlinear transport PDEs relevant to CFD, including nondimensional Burgers’ equation, by casting the governing equations and boundary conditions into an optimization problem over parametrized quantum circuits \cite{bengoechea2025toward}. Lubasch et al. introduce a “quantum nonlinear processing unit” (QNPU) that, combined with variational circuits, can approximate nonlinear PDEs such as the nonlinear Schrödinger and Burgers equations \cite{lubasch2020variational}. Pool et al. propose spacetime variational quantum algorithms (SVQAs) in which both space and time are encoded in qubit registers and the entire time evolution is obtained as the ground state of a Feynman--Kitaev-type Hamiltonian, demonstrating accurate Burgers solutions on IBM and Quantinuum devices \cite{pool2024nonlinear}. A related QCE implementation extends this spacetime formalism to general classical PDEs, including diffusive and turbulent transport, and shows that nonlinear effects can be captured within a single, global optimization over a spacetime-encoded ansatz \cite{pool2022solving}. Complementary to circuit-based approaches, hybrid quantum physics-informed neural networks (QPINNs) embed quantum subroutines into PINN architectures and have been shown to solve nonlinear PDEs such as Burgers and Navier--Stokes-type flows while targeting NISQ-friendly ansätze \cite{steijl2022quantum,berger2025trainable}. Collectively, these variational and PINN-based strategies underscore that robust PDE solving on today’s devices requires not only asymptotic algorithmic speedups but also hardware-aware optimizations and noise-tolerant architectures. 

At the level of CFD applications, Burgers-type systems are now used as realistic benchmarks for quantum-enhanced fluid simulation. Don Bosco et al. study the scalability and accuracy of a VQLS-based solver for a transient, two-dimensional, incompressible coupled Burgers system, comparing quantum solutions with GMRES and documenting resource requirements up to billions of effective mesh points in a hybrid workflow \cite{bosco2024demonstration}. Esmaeilifar et al. propose a pure quantum algorithm for a nonlinear Burgers equation aimed at high-speed compressible flows, addressing challenges posed by nonlinearity, no-cloning, and non-unitarity via multiple copies of the state vector, block-encoding, quantum Hadamard products, and linear combinations of unitaries, while reusing qubits across time steps to reduce resource scaling \cite{esmaeilifar2024quantum}. Uchida et al. exploit the Cole--Hopf transformation to simulate Burgers turbulence on a fault-tolerant quantum computer and focus on efficient extraction of stochastic quantities such as multi-point correlation functions from the quantum state, arguing that evaluation of statistical observables can exhibit exponential advantages in the number of spatial grid points compared to classical finite-difference methods \cite{uchida2024quantum}. 
Quantum differential-equation formulations of Burgers-like equations in Hilbert space have likewise been used to linearize nonlinear sonic processes and regularize turbulence-like phenomena in acoustics \cite{genccoglu2021use}. These developments sit within a broader effort to formulate quantum computation of fluid dynamics (QCFD). Recent reviews survey quantum algorithms for linear and nonlinear PDEs relevant to fluid mechanics, including Poisson, advection--diffusion, linearized Navier--Stokes, collisionless Boltzmann, and Burgers equations, and highlight practical bottlenecks such as state preparation, boundary-condition encoding, and measurement overhead \cite{amaral2025quantum,tennie2025quantum,gonzalez2025quantum}. Hydrodynamic Schrödinger equation (HSE) approaches and hybrid quantum--classical split-step solvers have also been tested on 1D Burgers-like flows, reinforcing the role of Burgers as a central benchmark for quantum CFD \cite{gonzalez2025quantum}. 

On current NISQ devices, hybrid variational and error-mitigation strategies have become the dominant paradigm. Foundational reviews of NISQ algorithms provide a comprehensive taxonomy of variational quantum eigensolvers, quantum approximate optimization algorithms, and other parameterized-circuit methods, detailing their objective functions, ansatz architectures, and optimization strategies as well as their limitations due to barren plateaus, expressibility issues, and hardware noise \cite{bharti2022noisy}. Complementary surveys of variational quantum algorithms highlight their role as noise-tolerant, shallow-depth approaches that make pragmatic use of limited coherence times by offloading optimization to classical processors \cite{benamer2025variational}. In parallel, a rapidly growing literature on quantum error correction (QEC) and quantum error mitigation (QEM) clarifies the trade-offs between full fault-tolerance and lighter-weight mitigation schemes. Devitt et al. review the fundamentals of QEC and fault-tolerant architectures, emphasizing the substantial overheads in qubits and gates required for large-scale algorithms \cite{devitt2013quantum}. Cai et al. provide a comprehensive review of QEM techniques, including zero-noise extrapolation, probabilistic error cancellation, symmetry and purity constraints, and learning-based mitigation, and quantify their sampling overheads and applicability \cite{cai2023quantum}. These mitigation techniques seek to bridge the gap between current NISQ hardware and full QEC, but they are typically algorithm-agnostic and do not exploit the specific structure of the underlying PDE or CFD problem.

Quantum machine learning (QML) and quantum neural networks (QNNs) provide another crucial layer in this landscape. Large-scale surveys of QML emphasize that hybrid quantum--classical models are the most realistic route to near-term impact, given hardware noise, limited qubit counts, and the cost of quantum random access memory \cite{lamichhane2025quantum,cerezo2022challenges}. 
They document applications of QML across cybersecurity, finance, healthcare, and drug discovery, and stress that carefully designed hybrid architectures can offer advantages in expressive power or parameter efficiency while remaining compatible with NISQ devices \cite{nguyen2024machine}. At the same time, detailed studies of QML generalization theory in the NISQ era point out that most existing analyses assume ideal noise-free quantum computers, whereas realistic devices introduce decoherence and sampling noise that can fundamentally affect generalization error bounds \cite{khanal2024generalization}. Systematic mappings of supervised QML under NISQ noise identify only a small set of works that rigorously quantify generalization behaviour and call for problem-specific benchmarks and tighter theoretical guarantees. More broadly, modern reviews of QML and quantum artificial intelligence (QAI) clarify the interplay between quantum data, hybrid architectures, and application-specific constraints, and argue that fully exploiting QML will require close integration of algorithm design, hardware-aware error mitigation, and classical post-processing or surrogate modelling \cite{siddi2025quantum,cai2023quantum}. 

Graph-structured learning is particularly natural in both PDE and circuit contexts. Classical CFD communities routinely exploit graph-based discretizations (finite-volume stencils, unstructured meshes), while quantum-computing communities have developed a growing body of work on quantum graph learning and quantum graph neural networks (QGNNs) \cite{shayeganfar2025quantum}. A tutorial by Verdon et al. on quantum graph recurrent neural networks (QGRNNs) illustrates how graph topology, local update rules, and recurrent dynamics can be encoded into quantum models \cite{choi2021tutorial}. Shayeganfar et al. systematically develop the taxonomy of quantum graph learning, including quantum random walks, quantum graph kernels, graph-based encoding schemes, and variational QGNN architectures for applications such as materials discovery and image classification \cite{shayeganfar2025quantum}. These studies show that graph-theoretic structures map efficiently onto quantum circuits and that QGNNs can reduce the number of trainable parameters relative to classical GNNs while achieving fast convergence on certain tasks, but they also emphasize significant challenges in encoding large graphs, handling complex noise models, and integrating QGNNs into end-to-end workflows for scientific computing. 

Despite this rapidly expanding ecosystem, there remains a notable gap at the intersection of QCFD, NISQ-era error mitigation, and graph-based learning. Most quantum Burgers and QCFD studies treat machine learning either as an alternative solver (e.g., QPINNs, QNN-based PDE solvers) or as an independent QML task, rather than as a learned, problem-aware correction layer that directly maps noisy quantum outputs to high-fidelity classical solutions. Existing QEM methods are typically circuit-level and do not exploit the spatial structure of the PDE discretization, while QML generalization studies rarely focus on physically structured regression tasks arising from CFD. Likewise, surveys of QML and QML-for-CFD underscore the promise of hybrid strategies but do not yet provide concrete, large-scale demonstrations of data-driven, graph-based error correction for quantum PDE solvers. These observations motivate the present work. We consider a hybrid quantum--classical pipeline for the one-dimensional viscous Burgers equation based on a Trotterized quantum solver that encodes the Cole--Hopf-transformed state on an $n$-qubit register and evolves it with nearest-neighbour entangling gates, in line with recent Hamiltonian-simulation-based quantum Burgers solvers \cite{oz2022solving}. 

For a wide parameter sweep over viscosity $\nu \in \{0.01, 0.05, 0.1, 0.15, 0.2\}$, time step $\Delta t \in \{5\times 10^{-4}, 10^{-3}, 1.5\times 10^{-3}, 2\times 10^{-3}\}$, grid size $N \in \{8, 16, 32, 64\}$, and left boundary velocity value $u_L \in \{1, 2, 4, 6\}$, we generate large ensembles of trajectories using both noisy quantum simulations (and IBM NISQ hardware) and a high-order classical reference solver. Building on recent advances in quantum Burgers algorithms, QCFD, NISQ variational methods, and QEM, we then construct a classical graph neural network whose nodes encode spatial grid points and whose features capture local flow variables, circuit-derived quantities, and coarse-grained diagnostics (e.g., shock position and dissipation rate). This GNN is trained to predict the reference classical solution from the noisy quantum outputs, effectively learning a low-cost surrogate error corrector that generalizes across $(\nu, \Delta t, N, u_L)$ and over time. Our contributions are threefold. First, we synthesize and extend recent developments in quantum Burgers solvers, QCFD, QML, and QEM into a unified perspective on quantum-enhanced nonlinear transport, with Burgers’ equation serving as a structurally rich yet tractable benchmark. Second, we present what is, to our knowledge, the first systematic study of a graph neural network trained as an error-correction layer for a quantum Burgers solver, leveraging large-scale parameter sweeps of noisy quantum circuits and classical CFD baselines in a NISQ-relevant setting. Finally, we demonstrate empirically that the learned GNN corrector significantly reduces solution error relative to raw NISQ outputs and simple zero-noise-extrapolation baselines, while preserving the qualitative structure of shocks and dissipation. 
In doing so, we illustrate how graph-based learning, quantum PDE algorithms, and modern error-mitigation concepts can be combined into a practical, scalable workflow for error-corrected quantum fluid simulation in the NISQ era. 

\section{\label{sec:level1}Problem Formulation and Classical Reference Solver}

In this section, we summarize the mathematical formulation of the viscous Burgers equation considered in this work, the Cole--Hopf transform used to linearize the problem, the spatial and temporal discretization of the resulting diffusion equation, and the diagnostic quantities that define our classical reference solution against which all quantum and learned error-corrected solutions are compared.

\subsection{Viscous Burgers Equation and Cole--Hopf Transform}
\label{subsec:burgers_cole_hopf}

We consider the one-dimensional viscous Burgers equation on a finite interval,
\begin{equation}
  \frac{\partial u}{\partial t}(x,t)
  + u(x,t)\,\frac{\partial u}{\partial x}(x,t)
  = \nu\,\frac{\partial^2 u}{\partial x^2}(x,t),
  \qquad x \in [0,1], \quad t \ge 0,
  \label{eq:burgers}
\end{equation}
where $u(x,t)$ is the (dimensionless) velocity field and $\nu>0$ is the viscosity.  
Throughout this work, we impose Dirichlet boundary conditions of the form:
\begin{equation}
  u(0,t) = u_L, 
  \qquad
  u(1,t) = u_R,
  \qquad t \ge 0,
  \label{eq:bc}
\end{equation}
and an initial condition:
\begin{equation}
  u(x,0) = u_0(x),
  \qquad x \in [0,1].
  \label{eq:ic}
\end{equation}
In our numerical experiments, we typically choose a smooth initial condition such as:
\begin{equation}
  u_0(x) = \sin(\pi x),
  \label{eq:ic_sine}
\end{equation}
which evolves under \eqref{eq:burgers} to form steep gradients and shock-like structures, while the boundary value $u_L$ is varied over a range of inflow conditions and $u_R$ is kept fixed (e.g., $u_R = 0$) in the parameter sweeps described later.

To obtain a linear evolution problem amenable to both classical and quantum solvers, we apply the Cole--Hopf transform.  
Defining a scalar potential $\phi(x,t)$ by:
\begin{equation}
  u(x,t) = -\,2\nu\,\frac{\partial}{\partial x}
  \bigl[ \ln \phi(x,t) \bigr]
  = -\,2\nu\,\frac{\phi_x(x,t)}{\phi(x,t)}.
  \label{eq:cole_hopf_relation}
\end{equation}
Substituting \eqref{eq:cole_hopf_relation} into \eqref{eq:burgers} and simplifying shows that $\phi(x,t)$ satisfies the linear diffusion (heat) equation:
\begin{equation}
  \frac{\partial \phi}{\partial t}(x,t)
  = \nu\,\frac{\partial^2 \phi}{\partial x^2}(x,t),
  \qquad x \in [0,1], \quad t \ge 0.
  \label{eq:heat}
\end{equation}
The initial condition for $\phi$ is obtained by integrating \eqref{eq:cole_hopf_relation} at $t=0$:
\begin{equation}
  \phi(x,0) \equiv \phi_0(x)
  = \exp\!\left(-\frac{1}{2\nu}\int_{0}^{x} u_0(y)\,\mathrm{d}y \right),
  \label{eq:phi0}
\end{equation}
up to a multiplicative normalization constant that does not affect $u$ in \eqref{eq:cole_hopf_relation}.  
In practice, we construct $\phi_0$ numerically from $u_0$ as described below.

Once \eqref{eq:heat} is solved for $\phi(x,t)$, the velocity field $u(x,t)$ is reconstructed via the discrete analogue of \eqref{eq:cole_hopf_relation}.  
This structure will be mirrored by our quantum solver: the quantum register encodes a discretized version of $\phi$, evolves it according to a diffusion generator, and we reconstruct $u$ via a discrete derivative of the amplitudes.

\subsection{Spatial Discretization and Boundary Conditions}
\label{subsec:spatial_disc}

We discretize the spatial domain $[0,1]$ on a uniform grid of $N$ points,
\begin{equation}
  x_j = j\,\Delta x,
  \qquad j = 0,1,\dots,N-1,
  \qquad \Delta x = \frac{1}{N-1}.
  \label{eq:grid}
\end{equation}
Let,
\begin{equation}
  \phi_j(t) \approx \phi(x_j,t), 
  \qquad
  u_j(t) \approx u(x_j,t),
\end{equation}
denote the discrete values of the Cole--Hopf field and velocity, respectively.

The second derivative in \eqref{eq:heat} is approximated by the standard second-order central-difference stencil,
\begin{equation}
  \frac{\partial^2 \phi}{\partial x^2}(x_j,t)
  \;\approx\;
  \frac{\phi_{j-1}(t) - 2\phi_j(t) + \phi_{j+1}(t)}{\Delta x^2},
  \qquad j = 1, \dots, N-2.
\end{equation}
This leads to the semi-discrete system:
\begin{equation}
  \frac{\mathrm{d}\boldsymbol{\phi}}{\mathrm{d}t}(t)
  = \nu\,L\,\boldsymbol{\phi}(t),
  \label{eq:semi_discrete}
\end{equation}
where $\boldsymbol{\phi}(t) = \bigl(\phi_0(t),\dots,\phi_{N-1}(t)\bigr)^{\mathsf{T}}$ and $L \in \mathbb{R}^{N\times N}$ is the discrete Laplacian matrix with entries:
\begin{equation}
  L_{j,j} = -\frac{2}{\Delta x^2}, \quad
  L_{j,j-1} = L_{j,j+1} = \frac{1}{\Delta x^2}, \quad
  j = 1,\dots,N-2,
  \label{eq:L_interior}
\end{equation}
and all other entries initially set to zero.  

We enforce Dirichlet boundary conditions at the level of $u$ rather than $\phi$: the boundary velocities are fixed to,
\begin{equation}
  u_0(t) = u_L, 
  \qquad 
  u_{N-1}(t) = u_R,
  \qquad \forall\,t \ge 0,
  \label{eq:disc_bc}
\end{equation}
while the corresponding boundary values of $\phi$ enter only implicitly through the reconstruction formula \eqref{eq:cole_hopf_relation}.  
Consistent with this choice, we zero out the first and last rows of $L$,
\begin{equation}
  L_{0,j} = 0, \qquad L_{N-1,j} = 0, \qquad \forall\, j,
\end{equation}
so that the interior diffusion dynamics evolve independently of any explicit boundary evolution for $\phi_0$ and $\phi_{N-1}$.  
In practice, this means that \eqref{eq:semi_discrete} is integrated for the interior entries, and the boundary velocities \eqref{eq:disc_bc} are imposed when reconstructing $u$ from $\boldsymbol{\phi}$.

The initial condition for $\boldsymbol{\phi}$ is constructed from the discrete version of \eqref{eq:phi0}.  
We first evaluate $u_0(x_j)$ from \eqref{eq:ic_sine}, then approximate the cumulative integral using the trapezoidal rule,
\begin{equation}
  I_j \approx \int_{0}^{x_j} u_0(y)\,\mathrm{d}y
  \;\approx\; \sum_{k=1}^{j}
  \frac{\Delta x}{2}\,\bigl(u_0(x_{k}) + u_0(x_{k-1})\bigr),
\end{equation}
and finally set:
\begin{equation}
  \phi_j(0) = \exp\!\left( -\frac{I_j}{2\nu} \right),
  \qquad j = 0,\dots,N-1.
  \label{eq:phi0_disc}
\end{equation}
A global normalization factor is applied so that $\boldsymbol{\phi}(0)$ has unit Euclidean norm; this convention is convenient for the quantum representation and does not affect the reconstructed $u$ through \eqref{eq:cole_hopf_relation}.

\subsection{Classical Time Integration and Krylov Solver}
\label{subsec:krylov_solver}

The semi-discrete evolution \eqref{eq:semi_discrete} admits the formal solution:
\begin{equation}
  \boldsymbol{\phi}(t) = \exp\!\bigl(\nu t L\bigr)\,\boldsymbol{\phi}(0).
  \label{eq:phi_solution}
\end{equation}
We evaluate the action of the matrix exponential on $\boldsymbol{\phi}(0)$ using a Krylov-subspace projection method.  
In particular, we treat \eqref{eq:semi_discrete} as a linear ODE system and employ a Krylov-based propagator (implemented via the krylovsolve routine in QuTiP) to approximate $\boldsymbol{\phi}(t)$ at a set of discrete times $t_n$.

Given a final time $T$ and a time step $\Delta t$, we define the temporal grid:
\begin{equation}
  t_n = n\,\Delta t, 
  \qquad n = 0,1,\dots,N_t,
  \qquad T = N_t \Delta t.
\end{equation}
For each target time $t_n$ that appears in the parameter sweeps and hardware experiments (see later sections), we construct a time list:
\begin{equation}
  \mathcal{T}_n = \{ 0, t_n \},
\end{equation}
and call the Krylov solver to obtain an approximation to $\boldsymbol{\phi}(t_n)$.  
Because $L$ is a sparse, banded matrix with real entries, the Krylov method is efficient and stable for the parameter ranges considered in this study. Once $\boldsymbol{\phi}(t_n)$ is available, the velocity field $\boldsymbol{u}(t_n)$ is reconstructed at the grid points using a discrete version of \eqref{eq:cole_hopf_relation}.  
For interior points $j = 1,\dots,N-2$, we use a centred difference,
\begin{equation}
  u_j(t_n)
  = -\,2\nu\,
  \frac{\phi_{j+1}(t_n) - \phi_{j-1}(t_n)}
       {2\,\Delta x\,\max\bigl(\phi_j(t_n),\varepsilon\bigr)},
  \label{eq:u_reconstruct}
\end{equation}
where $\varepsilon$ is a small positive regularization parameter (e.g.\ $\varepsilon = 10^{-10}$) that prevents division by extremely small values of $\phi_j$.  
The boundary values $u_0(t_n)$ and $u_{N-1}(t_n)$ are then set explicitly according to \eqref{eq:disc_bc}.  

The resulting discrete field:
\begin{equation}
  \mathbf{u}^{\mathrm{class}}(t_n)
  = \bigl(u_0(t_n),\dots,u_{N-1}(t_n)\bigr)^{\mathsf{T}}
\end{equation}
is taken as the classical reference solution at time $t_n$.  
All quantum and learned error-corrected solutions are benchmarked against this reference in the diagnostics defined below.

\subsection{Diagnostic Quantities: Errors, Shock Position, and Dissipation}
\label{subsec:diagnostics}

To quantify the agreement between the classical reference solver and quantum (or corrected) solutions, we define several diagnostic quantities.

Let $\mathbf{u}^{\mathrm{ref}}(t_n)$ denote the classical reference solution at time $t_n$ and $\mathbf{u}^{\mathrm{q}}(t_n)$ denote a quantum-derived approximation (e.g.\ from a noisy circuit, a zero-noise-extrapolated solution, or a GNN-corrected field).  
The discrete $L^2$-error at time $t_n$ is defined as:
\begin{equation}
  E_{L^2}(t_n)
  = \left(
      \Delta x \sum_{j=0}^{N-1}
      \bigl[ u^{\mathrm{q}}_j(t_n) - u^{\mathrm{ref}}_j(t_n) \bigr]^2
    \right)^{1/2}.
  \label{eq:L2_error}
\end{equation}
This quantity will be used to assess overall accuracy and to compare different mitigation strategies.

The location of the dominant shock (or steepest gradient) is estimated from the discrete spatial derivative.  
Defining the discrete gradient:
\begin{equation}
  \left(\frac{\partial u}{\partial x}\right)_{j+1/2}(t_n)
  \approx \frac{u_{j+1}(t_n) - u_{j}(t_n)}{\Delta x},
  \qquad j = 0,\dots,N-2,
  \label{eq:du_dx}
\end{equation}
and compute its magnitude.  
We define the shock position $x_{\mathrm{shock}}(t_n)$ as:
\begin{equation}
  x_{\mathrm{shock}}(t_n)
  = x_{k^\star}, \quad
  k^\star = \underset{0 \le j \le N-2}{\arg\max}\,
  \left|
    \frac{u_{j+1}(t_n) - u_{j}(t_n)}{\Delta x}
  \right|.
  \label{eq:shock_pos}
\end{equation}
This simple diagnostic is robust for the parameter regimes considered and allows us to track the propagation and sharpening of shock-like structures in time.

Finally, we define a discrete dissipation rate associated with the viscous term.  
In the continuous setting, the viscous dissipation density is proportional to $\nu \left(\partial u/\partial x\right)^2$.  
We approximate the total dissipation rate at time $t_n$ by:
\begin{equation}
  D(t_n)
  = \nu \sum_{j=0}^{N-2}
    \left[
      \left(\frac{u_{j+1}(t_n) - u_{j}(t_n)}{\Delta x}\right)^2
    \right]\Delta x.
  \label{eq:dissipation}
\end{equation}
Both $x_{\mathrm{shock}}(t_n)$ and $D(t_n)$ are computed for the classical reference solution and for the quantum/learned solutions, and their agreement (or discrepancy) provides a physically interpretable assessment of whether sharp gradients and energy dissipation are correctly captured.

The triple $(E_{L^2}(t_n), x_{\mathrm{shock}}(t_n), D(t_n))$ thus forms the main set of diagnostic quantities used throughout the remainder of the paper to evaluate and compare the quantum Burgers solver and the graph neural network error-correction model against the classical reference.

\section{\label{sec:level1}Quantum Burgers Solver on NISQ Hardware}

In this section, we describe the quantum solver used to approximate the Cole--Hopf–transformed Burgers dynamics, its implementation on noisy quantum hardware, and the baseline error-mitigation strategy employed to reduce hardware-induced errors. 
The workflow mirrors the classical formulation: we encode a discretized version of the Cole--Hopf field $\phi(x,t)$ into the amplitudes of an $n$-qubit register, apply a Trotterized diffusion operator built from nearest-neighbour $R_{XX}$ gates, measure the register in the computational basis, and reconstruct the velocity field $u(x,t)$ from the measurement statistics.

\subsection{Encoding of the Cole--Hopf State on Qubits}
\label{subsec:encoding}

We work on the same uniform spatial grid as in previous section, 
\begin{equation}
  x_j = j\,\Delta x, 
  \qquad j = 0,\dots,N-1, 
  \qquad \Delta x = \frac{1}{N-1},
\end{equation}
and use the Cole--Hopf transform to define the discrete initial field:
\begin{equation}
  \phi_j(0) = \exp\!\left( -\frac{I_j}{2\nu} \right),
  \qquad 
  I_j \approx \int_0^{x_j} u_0(y)\,\mathrm{d}y,
\end{equation}
where $u_0(x) = \sin(\pi x)$ is the initial velocity profile and $I_j$ is computed via the trapezoidal rule, cf.~\eqref{eq:phi0_disc}.  
We collect these values into a vector:
\begin{equation}
  \boldsymbol{\phi}_0 = 
  \bigl(\phi_0(0),\dots,\phi_{N-1}(0)\bigr)^{\mathsf{T}} \in \mathbb{R}^N,
\end{equation}
and normalize it to unit Euclidean norm,
\begin{equation}
  \tilde{\boldsymbol{\phi}}_0 
  = \frac{\boldsymbol{\phi}_0}{\|\boldsymbol{\phi}_0\|_2}.
\end{equation}
This normalization fixes an overall amplitude scale that is convenient for quantum state preparation; it does not affect the reconstructed velocity field $u$ obtained from amplitude ratios.

The number of qubits required to represent $N$ spatial points is:
\begin{equation}
  n = \left\lceil \log_2 N \right\rceil, 
  \qquad D = 2^n,
\end{equation}
and we embed $\tilde{\boldsymbol{\phi}}_0$ into a $D$-dimensional state by padding or truncation:
\begin{equation}
  \psi^{(0)}_k =
  \begin{cases}
    \tilde{\phi}_{k}, & 0 \le k \le N-1, \\
    0,                & N \le k \le D-1.
  \end{cases}
  \label{eq:psi0_embedding}
\end{equation}
A final renormalization step ensures,
\begin{equation}
  \sum_{k=0}^{D-1} \bigl|\psi^{(0)}_k\bigr|^2 = 1.
\end{equation}

We associate each computational basis state $\lvert k \rangle$ of the $n$-qubit register with the binary representation of the index $k$ and interpret $k$ as the spatial index $j$ for $0 \le j \le N-1$.  
The quantum register is then initialized in the pure state:
\begin{equation}
  \lvert \psi(0)\rangle
  = \sum_{k=0}^{D-1} \psi^{(0)}_k \,\lvert k \rangle,
\end{equation}
using the initialize instruction in Qiskit with the statevector $\psi^{(0)}$ from \eqref{eq:psi0_embedding}.  
This amplitude encoding ensures that, at $t=0$, the amplitudes on basis states $\lvert 0\rangle,\dots,\lvert N-1\rangle$ approximate the Cole--Hopf field $\phi(x_j,0)$ up to normalization.

\subsection{Trotterized \texorpdfstring{$R_{XX}$}{RXX} Time Evolution Circuit}
\label{subsec:trotter_circuit}

The classical semi-discrete evolution \eqref{eq:semi_discrete} reads:
\begin{equation}
  \frac{\mathrm{d}\boldsymbol{\phi}}{\mathrm{d}t}(t)
  = \nu\,L\,\boldsymbol{\phi}(t),
\end{equation}
with $L$ as the discrete Laplacian.  
To construct a quantum circuit that mimics this dynamics, we consider a nearest-neighbour Hamiltonian acting on the $n$-qubit register,
\begin{equation}
  H
  = \sum_{i=0}^{n-2} \alpha\,X_i X_{i+1},
  \label{eq:H_xx}
\end{equation}
where $X_i$ denotes the Pauli-$X$ operator on qubit $i$ and $\alpha$ is a scaling chosen such that the induced mixing between neighbouring amplitudes reproduces the discrete diffusion rate in the Cole--Hopf picture.  
In practice, this is realized using the two-qubit rotation gate $R_{XX}(\theta)$,
\begin{equation}
  R_{XX}(\theta) = \exp\!\left(-i \frac{\theta}{2} X \otimes X\right),
\end{equation}
applied to adjacent qubit pairs.

For a target physical time $t$ and a chosen time step $\Delta t$, the circuit is built from a first-order Trotter product,
\begin{equation}
  U(t) 
  \approx \bigl[U(\Delta t)\bigr]^{M},
  \qquad U(\Delta t) = \prod_{i=0}^{n-2} R_{XX}(\theta),
  \qquad M = \max\!\left(1, \left\lfloor \frac{t}{\Delta t} \right\rfloor \right).
  \label{eq:trotter_prod}
\end{equation}
We choose the rotation angle:
\begin{equation}
  \theta = \frac{2\nu\,\Delta t}{\Delta x^2},
  \label{eq:theta_choice}
\end{equation}
which matches the scale of the classical diffusion operator in \eqref{eq:semi_discrete}.  

Algorithmically, the quantum Burgers evolution circuit is constructed as follows:
\begin{enumerate}
  \item Allocate an $n$-qubit register and an $n$-bit classical register.
  \item Initialize the qubit register in the state $\lvert \psi(0)\rangle$ using $\psi^{(0)}$ from \eqref{eq:psi0_embedding}.
  \item Compute the number of Trotter steps $M = \max(1,\lfloor t/\Delta t \rfloor)$.
  \item For each Trotter step $m = 1,\dots,M$:
    \begin{enumerate}
      \item For each nearest-neighbour pair $(i,i+1)$, apply $R_{XX}(\theta)$ between qubits $i$ and $i+1$.
    \end{enumerate}
  \item Insert a barrier to separate evolution and measurement.
  \item Measure all qubits in the computational basis and store the outcome bitstrings in the classical register.
\end{enumerate}  

Given a circuit for time $t$, we execute it with $N_{\mathrm{shots}}$ measurements (typically $N_{\mathrm{shots}} = 8192$) and obtain a count distribution $\texttt{counts}(b)$ over bitstrings $b \in \{0,1\}^n$.  
To reconstruct an approximate amplitude vector on the spatial grid, we proceed as:
\begin{equation}
  \psi_k(t) 
  = \frac{\sqrt{\texttt{counts}\bigl(b_k\bigr)}}{\sqrt{N_{\mathrm{shots}}}},
  \qquad b_k = \text{binary representation of }k,
  \label{eq:psi_from_counts}
\end{equation}
followed by normalization:
\begin{equation}
  \psi_k(t) \leftarrow 
  \frac{\psi_k(t)}{\left(\sum_{j=0}^{N-1} |\psi_j(t)|^2 \right)^{1/2}}.
\end{equation}
This amplitude vector plays the role of a discrete Cole--Hopf field at time $t$; the approximate velocity field is then reconstructed from $\psi(t)$ using the same discrete derivative formula as in the classical case,
\begin{equation}
  u_j^{\mathrm{q}}(t)
  = -\,2\nu\,
  \frac{\psi_{j+1}(t) - \psi_{j-1}(t)}
       {2\,\Delta x\,\max\bigl(\psi_j(t),\varepsilon\bigr)},
  \qquad j=1,\dots,N-2,
  \label{eq:uq_from_psi}
\end{equation}
with $u_0^{\mathrm{q}}(t) = u_L$ and $u_{N-1}^{\mathrm{q}}(t) = u_R$ enforced as boundary values.

\subsection{Simulation Backends and IBM Hardware Configuration}
\label{subsec:hardware_config}

We implement the quantum Burgers solver using Qiskit and Qiskit Aer.  
Two types of backends are employed:
\begin{itemize}
  \item A noisy simulator, based on AerSimulator, configured with a realistic noise model extracted from an IBM Quantum device.
  \item A real NISQ device, accessed through IBM Quantum Runtime.
\end{itemize}

For hardware experiments, we use the IBM backend and initialize the runtime service. We then attach a runtime Session to this backend and generate a device-aware transpilation pass manager which is used to transpile all circuits prior to execution.  

To emulate hardware noise in simulation, we construct a noise model from the same backend and pass it to the AerSimulator. Noisy simulations and hardware runs therefore experience comparable gate and readout error profiles. For each parameter setting $(\nu,\Delta t,N,u_L)$ and each target time $t$ in the selected time grids, we:
\begin{enumerate}
  \item Build the Trotterized evolution circuit as described in Sec.~\ref{subsec:trotter_circuit}.
  \item Transpile the circuit for the noisy simulator using the pass manager.
  \item Execute the transpiled circuit on the AerSimulator with $N_{\mathrm{shots}} = 8192$ and the calibrated noise model, obtaining noisy counts $\texttt{counts}_{\mathrm{sim}}$.
  \item Optionally transpile the same logical circuit for the hardware backend and submit it via the IBM Runtime Sampler API, obtaining either a quasi-probability distribution or counts $\texttt{counts}_{\mathrm{hw}}$.
\end{enumerate}
Both simulated and hardware counts are converted to amplitude vectors using \eqref{eq:psi_from_counts}, followed by reconstruction of the velocity field via \eqref{eq:uq_from_psi}.  

For reproducibility and downstream analysis, we export every circuit (noisy simulator, ZNE-augmented, and hardware-transpiled versions) to both QPY and QASM format, and we store the resulting metadata, diagnostic quantities, and reconstructed velocity fields in structured JSON files.  
These exports include the circuit depth, the total number of two-qubit gates, the backend configuration, and references to the stored QPY/QASM files.

\subsection{Zero-Noise Extrapolation as Baseline Error Mitigation}
\label{subsec:zne}

To obtain a hardware-agnostic baseline error-mitigated solution, we employ a simple form of zero-noise extrapolation (ZNE) on the noisy simulator.  
The idea is to systematically amplify the effective noise in the circuit, measure the observable of interest at several noise scales, and extrapolate back to the zero-noise limit.

In our implementation, noise amplification is achieved by repeating the entangling $R_{XX}$ gate in the Trotter block.  
Specifically, we define a scaled Trotter circuit:
\begin{equation}
  U_{\mathrm{scaled}}(t; s)
  \approx \left[
    \prod_{i=0}^{n-2} R_{XX}(\theta)^{s}
  \right]^{M},
\end{equation}
where $s \in \mathbb{N}$ is an integer noise scale factor and $M$ is the number of Trotter steps as in \eqref{eq:trotter_prod}.  
When $s=1$, we recover the nominal circuit; larger $s$ increase the circuit depth and thus the accumulated noise while keeping the ideal (noiseless) evolution unchanged up to global phases.

For each time $t$ of interest, we consider two noise scales,
\begin{equation}
  s \in \{1,3\},
\end{equation}
and construct the corresponding circuits using a dedicated routine.  
For each $s$:
\begin{enumerate}
  \item We build the scaled Trotter circuit $U_{\mathrm{scaled}}(t; s)$.
  \item We transpile it with the same device-aware pass manager.
  \item We execute it on the noisy simulator with the calibrated noise model and $N_{\mathrm{shots}} = 8192$.
  \item From the resulting counts, we reconstruct an amplitude vector $\psi^{(s)}(t)$ and a velocity field $u^{(s)}(t)$ using \eqref{eq:psi_from_counts} and \eqref{eq:uq_from_psi}.
\end{enumerate}

Assuming that the error in $u^{(s)}(t)$ depends approximately linearly on the noise scale $s$ for moderate $s$, we perform a first-order Richardson extrapolation to estimate the zero-noise limit:
\begin{equation}
  u^{\mathrm{ZNE}}(t)
  = \frac{3}{2}\,u^{(1)}(t) - \frac{1}{2}\,u^{(3)}(t).
  \label{eq:zne_u}
\end{equation}
The extrapolated field is then clipped to a physically reasonable range,
\begin{equation}
  u^{\mathrm{ZNE}}_j(t) \;\leftarrow\;
  \min\!\bigl(\max(u^{\mathrm{ZNE}}_j(t), u_{\min}), u_{\max}\bigr),
\end{equation}
with $u_{\min}$ and $u_{\max}$ chosen to bound the observed velocities (e.g.\ $u_{\min}=-1$, $u_{\max}=2$ in our experiments).

For each parameter setting and time $t$, we thus obtain three quantum-derived solutions:
\begin{itemize}
  \item A noisy solution $u^{\mathrm{noisy}}(t)$ from the nominal circuit ($s=1$).
  \item A ZNE-corrected solution $u^{\mathrm{ZNE}}(t)$ from \eqref{eq:zne_u}.
  \item When available, a hardware solution $u^{\mathrm{hw}}(t)$ from the IBM device.
\end{itemize}
All three are compared to the classical reference $u^{\mathrm{class}}(t)$ using the diagnostics in Sec.~\ref{subsec:diagnostics}.  
In later sections, these quantum solutions constitute the input to our graph neural network error corrector, which is trained to learn the residual mapping:
\begin{equation}
  \mathcal{R}:\;
  u^{\mathrm{noisy}}(t) 
  \;\text{or}\;
  u^{\mathrm{ZNE}}(t)
  \;\mapsto\;
  u^{\mathrm{class}}(t),
\end{equation}
and thereby provide a learned, problem-aware error-correction layer for the quantum Burgers solver.

\section{\label{sec:level1}Dataset Generation and Parameter Sweep}

This section describes how we generate a large, structured dataset of classical and quantum solutions of the viscous Burgers equation across a broad range of parameters.  
As can be seen in Fig. 1, each sample in the dataset consists of: (i) a classical reference solution obtained from the Cole--Hopf–based Krylov solver of Sec II, (ii) one or more quantum solutions obtained from the noisy Trotterized circuit of Sec III (with and without zero-noise extrapolation), and (iii) detailed circuit metadata (depth, gate counts, and file references) exported in QPY/QASM and JSON formats.  
These paired samples serve as training and evaluation data for the graph neural network (GNN) error corrector introduced later.

\begin{figure*}
    \centering
    \includegraphics[width=0.99\linewidth]{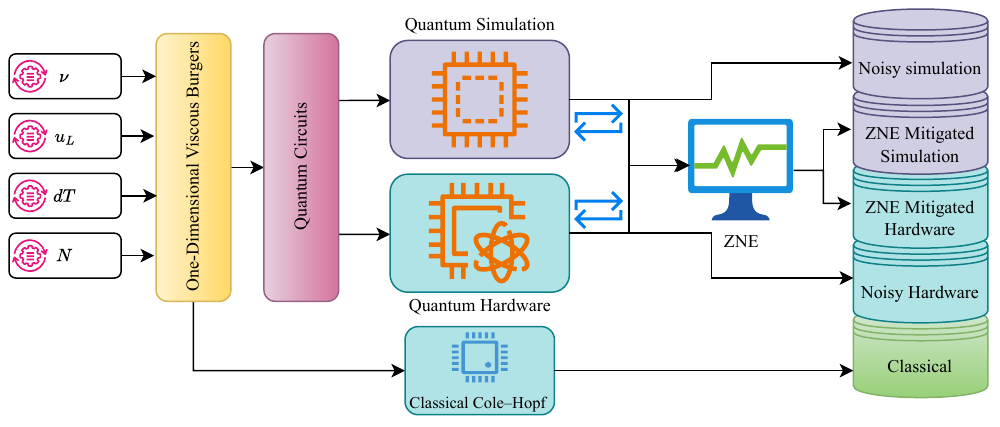}
    \caption{Data generation for classical/quantum simulations over classical and quantum hardwares of the 1D viscous Burger's equation.}
    \label{fig:placeholder}
\end{figure*}

\subsection{Parameter Space: \texorpdfstring{$\nu$}, $\Delta t$, $N$, and $u_L$}
\label{subsec:param_space}

To probe a representative set of viscous and numerical regimes, we perform a systematic parameter sweep over:
\begin{equation}
  \nu \in \{0.01, 0.05, 0.10, 0.15, 0.20\}
\end{equation}
\begin{equation}
  \Delta t \in \{5\times 10^{-4}, 10^{-3}, 1.5\times 10^{-3}, 2\times 10^{-3}\}
\end{equation}
\begin{equation}
  N \in \{8, 16, 32, 64\},
  \qquad
  u_L \in \{1, 2, 4, 6\},
\end{equation}
with $u_R = 0$ in all cases.  
The viscosity range spans moderately diffusive to relatively sharp-shock regimes; the time steps include values that are small enough to resolve the dynamics yet large enough to generate nontrivial circuit depths; and the grid sizes cover one to six qubits ($N=8$ corresponds to $n=3$, $N=64$ to $n=6$) after padding to the nearest power of two.  
In total, the grid contains 320 distinct parameter combinations.

For each parameter combination $(\nu,\Delta t, N, u_L)$, we evolve the system to a set of target times that are reused across classical, simulated-quantum, and hardware runs.  
We organize these times into three time sets:
\begin{align}
  \mathcal{T}^{(0)} &= \{ 0.0,\ 0.005,\ 0.010 \}, \\
  \mathcal{T}^{(1)} &= \{ 0.0,\ 0.0025,\ 0.005,\ 0.0075,\ 0.010 \}, \\
  \mathcal{T}^{(2)} &= \{ 0.0,\ 0.002,\ 0.004,\ 0.006,\ 0.008,\ 0.010 \}.
\end{align}
These sets provide increasingly refined temporal sampling near the early-time evolution where shock formation and steep gradients are most sensitive to viscosity and resolution. Each parameter combination thus yields 14 time snapshots, leading to a few thousand total samples across the sweep.

\subsection{Classical vs.~Quantum Runs and Measurement Protocol}
\label{subsec:classical_quantum_runs}

For each parameter setting $(\nu,\Delta t, N, u_L)$ and each time set $\mathcal{T}^{(s)}$, $s\in\{0,1,2\}$, we proceed in two phases: a classical phase and a quantum phase.

\subsubsection{Classical phase}

In the classical phase, we first reconstruct the Laplacian matrix $L \in \mathbb{R}^{N \times N}$ and Cole--Hopf initial state $\boldsymbol{\phi}(0)$ as in Sec II.  
We then form the linear operator:
\begin{equation}
  A = \nu\,L,
\end{equation}
and use a Krylov-based time propagator to evaluate:
\begin{equation}
  \boldsymbol{\phi}(t) = \exp(A t)\,\boldsymbol{\phi}(0)
\end{equation}
for each $t \in \mathcal{T}^{(s)}$.  

For each time $t$ in a given time set, we:
\begin{enumerate}
  \item Call the Krylov solver with a time list from $0$ to $t$ to obtain an approximation to $\boldsymbol{\phi}(t)$.
  \item Reconstruct the velocity field $\mathbf{u}^{\mathrm{class}}(t)$ using the discrete Cole--Hopf relation
    \begin{equation}
      u^{\mathrm{class}}_j(t)
      = -\,2\nu\,
      \frac{\phi_{j+1}(t) - \phi_{j-1}(t)}
           {2\,\Delta x\,\max\bigl(\phi_j(t),\varepsilon\bigr)},
      \qquad j = 1,\dots,N-2,
    \end{equation}
    with boundary values $u^{\mathrm{class}}_0(t)=u_L$, $u^{\mathrm{class}}_{N-1}(t)=u_R$.
  \item Compute the diagnostic quantities (shock position and dissipation) for the classical solution,
    \begin{equation}
      x_{\mathrm{shock}}^{\mathrm{class}}(t), 
      \qquad
      D^{\mathrm{class}}(t),
    \end{equation}
    using the definitions in Sec.~\ref{subsec:diagnostics}.
\end{enumerate}
The resulting classical solutions $\{\mathbf{u}^{\mathrm{class}}(t)\}$ for all times $t$ in all time sets are cached and later stored in the dataset as reference fields.

\subsubsection{Quantum phase}

In the quantum phase, we build the Trotterized evolution circuit that encodes the same initial state and time evolution, as detailed in Sec.~\ref{subsec:trotter_circuit}, and execute it on a noisy simulator and, when enabled, on an IBM NISQ device.

For each $(\nu,\Delta t, N, u_L)$ and $t \in \mathcal{T}^{(s)}$:
\begin{enumerate}
  \item We construct the $n$-qubit Trotter circuit:
    \begin{equation}
      U(t) \approx \bigl[U(\Delta t)\bigr]^M,
    \end{equation}
    with $M = \max(1,\lfloor t/\Delta t\rfloor)$ and $R_{XX}(\theta)$ rotations with:
    \begin{equation}
      \theta = \frac{2\nu\,\Delta t}{\Delta x^2}.
    \end{equation}
  \item We transpile $U(t)$ for a noisy simulator using the device-aware pass manager.
  \item We execute the transpiled circuit on AerSimulator with a noise model and $N_{\mathrm{shots}} = 8192$, obtaining a count distribution $\texttt{counts}_{\mathrm{sim}}(b)$.
  \item From the counts, we reconstruct a normalized amplitude vector:
    \begin{equation}
      \psi^{\mathrm{sim}}_j(t)
      = \frac{\sqrt{\texttt{counts}_{\mathrm{sim}}(b_j)}}{\sqrt{N_{\mathrm{shots}}}},
      \qquad j = 0,\dots,N-1,
    \end{equation}
    and the corresponding velocity field $\mathbf{u}^{\mathrm{noisy}}(t)$ via:
    \begin{equation}
      u^{\mathrm{noisy}}_j(t)
      = -\,2\nu\,
      \frac{\psi^{\mathrm{sim}}_{j+1}(t) - \psi^{\mathrm{sim}}_{j-1}(t)}
           {2\,\Delta x\,\max\bigl(\psi^{\mathrm{sim}}_j(t),\varepsilon\bigr)},
      \qquad j=1,\dots,N-2.
    \end{equation}
  \item We compute shock and dissipation diagnostics for the noisy solution,
    \begin{equation}
      x_{\mathrm{shock}}^{\mathrm{noisy}}(t), 
      \qquad D^{\mathrm{noisy}}(t),
    \end{equation}
    as in Sec.~\ref{subsec:diagnostics}.
\end{enumerate}

We then apply zero-noise extrapolation (ZNE) as described in Sec.~\ref{subsec:zne}.  
For each time $t$, we build scaled circuits with gate-repetition factors $s \in \{1,3\}$, run them on the same noisy simulator, reconstruct $\mathbf{u}^{(1)}(t)$ and $\mathbf{u}^{(3)}(t)$, and form the ZNE estimate:
\begin{equation}
  \mathbf{u}^{\mathrm{ZNE}}(t)
  = \frac{3}{2}\,\mathbf{u}^{(1)}(t)
  - \frac{1}{2}\,\mathbf{u}^{(3)}(t),
\end{equation}
with clipping to a physically reasonable range.  
This yields diagnostics $x_{\mathrm{shock}}^{\mathrm{ZNE}}(t)$ and $D^{\mathrm{ZNE}}(t)$ analogous to the classical and noisy cases.

When hardware execution is enabled, we additionally transpile $U(t)$ for an IBM backend and submit it via the IBM Runtime.  
The returned quasi-probability distribution or counts are converted to $\mathbf{u}^{\mathrm{hw}}(t)$ using the same reconstruction pipeline, and the corresponding diagnostics $x_{\mathrm{shock}}^{\mathrm{hw}}(t)$, $D^{\mathrm{hw}}(t)$ are computed.  
In practice, hardware runs may be carried out for a subset of the parameter grid due to queue and resource constraints; the dataset format accommodates both the presence and absence of hardware data.

\subsection{QPY/QASM, JSON Schema, and Metadata}
\label{subsec:export_pipeline}

To support downstream analysis, reproducibility, and reuse, we export all circuits and numerical results into a structured on-disk layout.  
For each parameter combination $(\nu,\Delta t, N, u_L)$, we create a unique experiment tag and an associated JSON file in the directory. For each time set index $s$ and each time $t \in \mathcal{T}^{(s)}$, we define a base stem. From this base stem, we generate three circuit files:
\begin{itemize}
  \item A noisy-simulator circuit,
  \item A ZNE-simulator circuit,
  \item When applicable, a hardware-transpiled circuit.
\end{itemize}
Each circuit is saved both as a QPY file and, when supported, a QASM file. A helper routine returns a dictionary with the circuit name, QPY path, and QASM path, which is stored in the JSON metadata. The top-level JSON file for each experiment contains:
\begin{itemize}
  \item A schema block (version, task description, and fields).
  \item The base name, timestamp, and output directory.
  \item Flags indicating whether hardware was used and, if so, which backend.
  \item The total number of time sets and records, and the total wall-clock time for the run.
  \item A list records[] containing one entry per time snapshot.
\end{itemize}
Each record includes:

\begin{itemize}
  \item \textbf{Grid and parameters:} \texttt{grid = \{N, dx\}}, \texttt{params = \{$\nu, u\_L, u\_R, dt$\} }.
  
  \item \textbf{Circuit references:} paths and names for the noisy and ZNE circuits (and hardware circuit when present).
  \item \textbf{Metrics:} circuit depth and two-qubit-gate counts for noisy and ZNE circuits; classical shock position and dissipation rate.
  \item \textbf{Outputs:}
    \begin{itemize}
      \item Classical: the classical reference field $\mathbf{u}^{\mathrm{class}}(t)$;
      \item Sim: the noisy field $\mathbf{u}^{\mathrm{noisy}}(t)$, the ZNE field $\mathbf{u}^{\mathrm{ZNE}}(t)$ (if computed), and raw noisy counts;
      \item Hardware: when present, the hardware field $\mathbf{u}^{\mathrm{hw}}(t)$, raw counts or quasi-distributions, and associated diagnostics.
    \end{itemize}
\end{itemize}

\subsection{Summary of Dataset Statistics and Regimes Covered}
\label{subsec:dataset_stats}

The full parameter sweep over $(\nu,\Delta t, N, u_L)$ yields $320$ distinct parameter combinations.  
For each combination we generate up to $14$ time snapshots across the three time sets $\mathcal{T}^{(0)}$, $\mathcal{T}^{(1)}$, and $\mathcal{T}^{(2)}$, resulting in several thousand paired samples of:
\begin{equation}
  \bigl(
    \mathbf{u}^{\mathrm{class}}(t),
    \mathbf{u}^{\mathrm{noisy}}(t),
    \mathbf{u}^{\mathrm{ZNE}}(t),
    \mathbf{u}^{\mathrm{hw}}(t) \text{ (when available)}
  \bigr),
\end{equation}
along with detailed circuit metadata and diagnostic quantities. The dataset spans:
\begin{itemize}
  \item \textbf{Viscosity regimes:} from $\nu=0.01$ (sharper shocks, stronger nonlinearity) to $\nu=0.2$ (more diffusive, smoother profiles);
  \item \textbf{Temporal regimes:} early-time evolution ($t \approx 0$) through $t = 0.01$, where shock formation and dissipation dynamics are most pronounced;
  \item \textbf{Spatial resolutions:} coarse grids ($N=8$) to moderately fine grids ($N=64$), mapping to $n=3$ to $n=6$ qubits in the quantum encoding;
  \item \textbf{Boundary conditions:} multiple inflow states via $u_L \in \{1,2,4,6\}$ with fixed outflow $u_R=0$.
\end{itemize}
From the perspective of learning, this collection provides a diverse set of input–output pairs, noisy and ZNE quantum fields together with high-fidelity classical references, covering a variety of shock strengths, dissipation rates, and grid resolutions. In the next section, we describe how these structured, graph-like datasets are used to train a graph neural network that acts as an error-correction layer for the quantum Burgers solver.

\section{\label{sec:level1} Attention Graph Neural Network-Based QEM}

Quantum error mitigation (QEM) has emerged as a critical technique for improving the reliability of quantum computations in the presence of noise. Traditional QEM approaches often rely on circuit-level or gate-level abstractions that may not fully capture the complex interdependencies within quantum circuits. More importantly, they are suffering from a high computational cost which prohibits their applicability in large-scale problems such as the one introduced in this work. Inspired by our previous work \cite{tousi2025qagtmLP}, we are using a novel graph-based approach that uses the structural properties of quantum circuits to enhance error mitigation performance.

\subsection{Graph Representation of Quantum Circuits}

Our method begins by representing quantum circuits as directed acyclic graphs (DAGs), where nodes correspond to quantum gates and edges represent the flow of quantum information between operations. This graph representation naturally captures the causal structure and dependencies inherent in quantum computations, providing a rich foundation for learning-based error mitigation. Each node in the circuit graph is associated with a feature vector that encodes relevant properties of the corresponding quantum gate, including gate type and parameters, Qubit indices and connectivity, temporal ordering within the circuit, and local noise characteristics. The graph structure enables our model to understand how errors propagate through the circuit and how different gates influence the final measurement outcomes. This representation is particularly well-suited for capturing the non-local correlations that arise in quantum systems due to entanglement, cross-talk and/or interference effects.

\subsection{Attention-Based Graph Neural Network Architecture}

As can be observed in Fig. 2, at the core of our QEM approach lies an attention-based graph neural network that processes the circuit representation to predict and correct quantum measurement errors. The architecture consists of several key components:

\begin{figure*}
    \centering
    \includegraphics[width=0.99\linewidth]{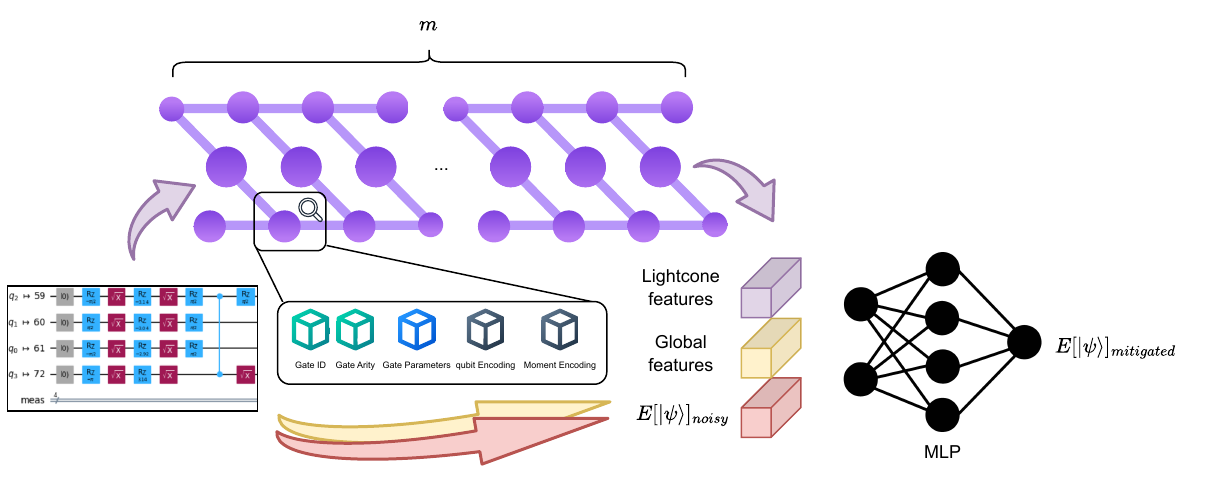}
    \caption{Quantum Error Mitigation (QEM) Graph Neural Network model schematic for training the noisy spatiotemporal circuits of 1D Burger's equation.}
    \label{fig:placeholder}
\end{figure*}

\subsubsection{Graph Encoder Layers}

The graph encoder employs multiple layers of graph attention networks (GATs) to learn representations of the circuit structure. Each attention layer computes node embeddings by aggregating information from neighboring nodes, with attention weights determined by the learned importance of different connections. This mechanism allows the model to focus on the most relevant parts of the circuit for error correction.

\paragraph{Lightcone Local Masks}
A crucial component of our graph embedding approach is the incorporation of lightcone local masks, which capture the causal influence structure within quantum circuits. The lightcone mask for each measured qubit defines the subset of gates that can causally affect the measurement outcome of that specific qubit. This concept is fundamental in quantum mechanics, where information propagation is constrained by the circuit's topology and the finite speed of quantum information transfer.

Formally, for a measured qubit $q_i$, the lightcone mask $L_i$ is defined as:
\begin{equation}
L_i = \{g \in G : \exists \text{ causal path from } g \text{ to measurement of } q_i\}
\end{equation}
where $G$ represents the set of all gates in the circuit. These masks serve multiple purposes:

\begin{itemize}
    \item \textbf{Causal Consistency}: Ensure that only causally relevant gates contribute to the prediction of each measurement outcome
    \item \textbf{Computational Efficiency}: Reduce the effective graph size by focusing attention on relevant substructures
    \item \textbf{Physical Interpretability}: Maintain alignment with the fundamental principles of quantum information theory
    \item \textbf{Noise Localization}: Enable the model to understand how local noise sources propagate through specific causal pathways
\end{itemize}

The lightcone masks are computed during the circuit-to-graph conversion process using topological analysis of the quantum circuit. Each gate node is assigned to the lightcones of all qubits that it can causally influence, creating a structured partitioning of the circuit graph that respects quantum mechanical causality.

\paragraph{General Node Features}
Each node in the circuit graph is characterized by a comprehensive feature vector that encodes both structural and physical properties of the corresponding quantum gate. This rich feature representation enables the graph neural network to learn nuanced relationships between circuit structure, gate properties, and measurement outcomes. The combination of lightcone-aware processing and comprehensive node features allows our model to capture both local gate-level effects and global circuit-level correlations that are essential for effective quantum error mitigation.

\subsubsection{Multi-Scale Processing}
Our architecture incorporates multi-scale processing to handle quantum circuits of varying sizes and complexities. The model can adapt to different output dimensions (e.g., 8, 16, 32, or 64 measurement outcomes) by dynamically adjusting its output layers while maintaining the same core processing pipeline.

\subsection{Training Methodology}

The training process involves learning to map from noisy quantum measurements to their corresponding noise-free counterparts. We employ a supervised learning approach where the model is trained on pairs of circuit graphs and their associated measurement data:

\begin{itemize}
    \item \textbf{Input}: Circuit graph representation in addition to noisy measurement outcomes
    \item \textbf{Target}: Classical simulation results (ground truth)
    \item \textbf{Loss Function}: Mean squared error between predicted and true outcomes
\end{itemize}

To enhance the model's robustness and generalization capabilities, we implement several training strategies:

\subsubsection{Incremental Hardware Integration}
We develop an incremental training approach that gradually substitutes simulated noisy data with real hardware measurements. This methodology allows us to study how the incorporation of actual quantum hardware noise affects model performance and to find optimal ratios of simulation versus hardware data for training.

\subsubsection{Dimension-Specific Models}
Rather than training a single model for all output dimensions, we train separate specialized models for each target dimension (16, 32, 64). This approach allows each model to optimize its parameters specifically for the characteristics and challenges associated with its target output size.

\subsection{Error mitigation Pipeline}

The complete error mitigation pipeline operates as follows:

\begin{enumerate}
    \item \textbf{Circuit Analysis}: Convert the quantum circuit into a graph representation with appropriate node features and edge connections.
    
    \item \textbf{Graph Processing}: Apply the graph neural network to encode the circuit structure and extract relevant features for error mitigation.
    
    \item \textbf{Error Prediction}: Use the multi-layer perceptron (MLP) component to predict the noise-corrected measurement outcomes based on the processed circuit information and noisy measurements.
    
\end{enumerate}

\subsection{Advantages of the Graph-Based Approach}

Our QAGT-MLP method offers several key advantages over traditional approaches:

\begin{itemize}
    \item \textbf{Structural Awareness}: By explicitly modeling circuit structure, the method can leverage topological information to improve error correction accuracy.
    
    \item \textbf{Scalability}: The graph representation naturally scales to circuits of different sizes and complexities without requiring architectural modifications.
    
    \item \textbf{Interpretability}: The attention mechanisms provide insights into which parts of the circuit are most important for error mitigation, enabling better understanding of the mitigation process.
    
    \item \textbf{Hardware Adaptability}: The method can be trained on both simulated and real hardware data, allowing it to adapt to the specific noise characteristics of different quantum devices.
    
    \item \textbf{Generalization}: The learned representations can potentially transfer across different circuit families and quantum algorithms, reducing the need for task-specific retraining.
\end{itemize}

This graph-based approach represents a significant advancement in quantum error mitigation, providing a principled framework for leveraging the structural properties of quantum circuits to achieve more effective noise correction. The following sections present detailed experimental results demonstrating the effectiveness of this method across various quantum computing scenarios.

\section{\label{sec:level1} Results and Discussion}

We now evaluate the graph neural network--based quantum error mitigation (QAGT-MLP) model introduced in the previous section.  
Unless otherwise stated, the QAGT-MLP model is trained on simulated noisy quantum data generated by the Trotterized Burgers solver and tested either (i) on held-out noisy simulation data or (ii) on real hardware outputs from the IBM backend, using the classical Cole--Hopf solver as ground truth in both cases.

\subsection{Baseline Training Behavior and Validation Performance}
\label{subsec:training_behavior}

\begin{figure*}[t]
  \centering
  \includegraphics[width=\textwidth]{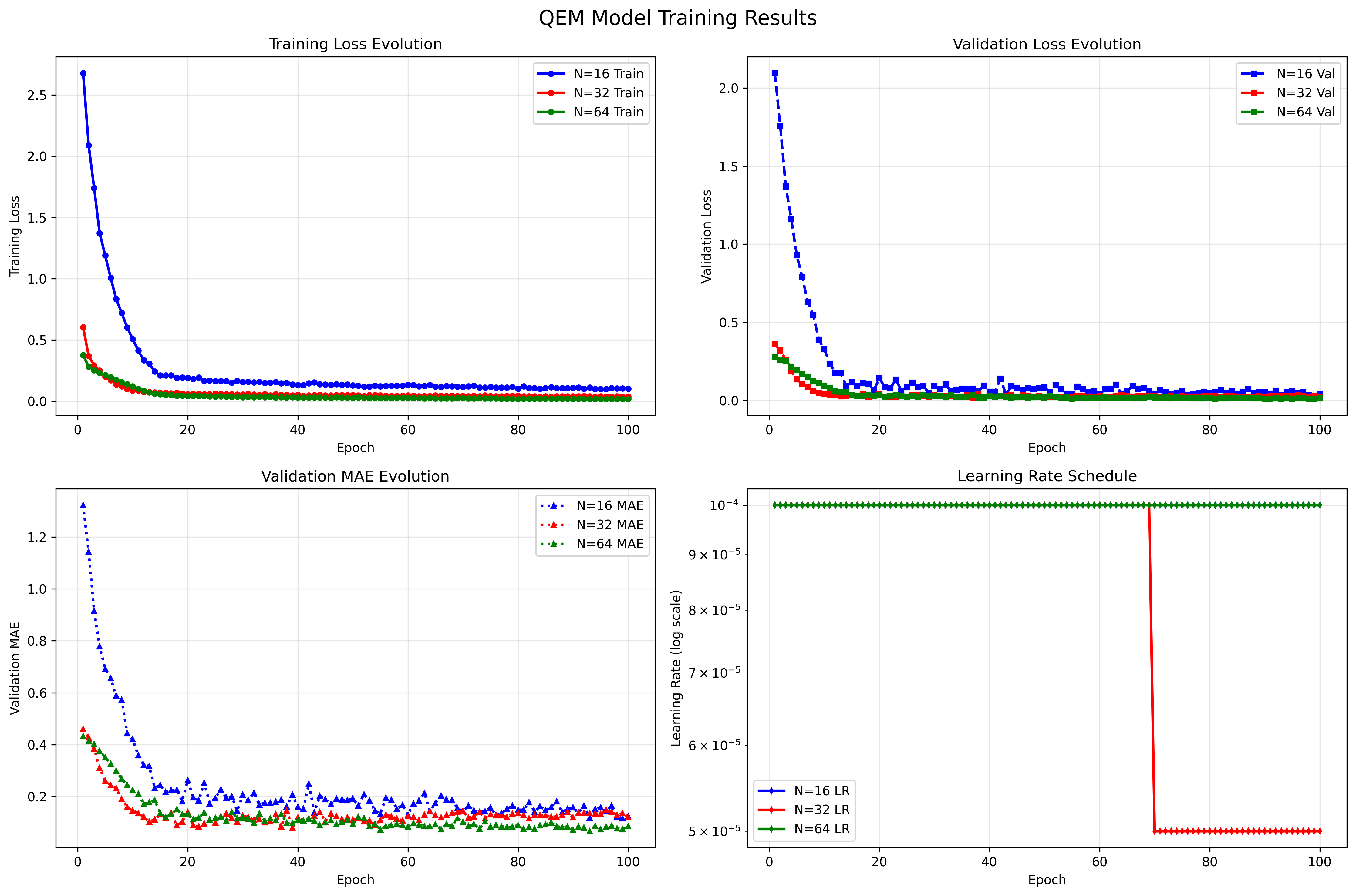}
  \caption{%
  Training and validation loss for
    $N=16,32,64$. %
  }
  \label{fig:training_curves1}
\end{figure*}

Fig.~\ref{fig:training_curves1} summarizes the training dynamics of the QAGT-MLP model for three spatial resolutions $N \in \{16, 32, 64\}$.  
The top row shows the evolution of the training and validation loss (mean-squared error) over $100$ epochs.  
For all three resolutions the training loss decreases rapidly during the first $10$--$15$ epochs and then gradually plateaus, while the validation loss tracks the training curve closely, indicating that the model does not exhibit strong overfitting within the epoch range considered.

The dependence on spatial resolution is mild as models trained at $N=32$ and $N=64$ start with significantly smaller initial loss than $N=16$, reflecting the fact that coarser grids produce noisier and more aliased inputs.  
By epoch $\sim 30$ all three resolutions reach a regime of slow, stable improvement with validation loss of $\mathcal{O}(10^{-2})$. The bottom-left panel of Fig.~\ref{fig:training_curves1} reports the evolution of the validation mean absolute error (MAE). The MAE curves mirror the loss behavior, with $N=64$ consistently yielding the smallest error and $N=16$ the largest. It also shows the learning-rate schedule used in training: a constant learning rate for $N=16$ and $N=64$, and a step decrease for $N=32$ around epoch $70$, which slightly sharpens the late-epoch convergence without inducing instability.

\subsection{Error Reduction on Simulated Quantum Data}
\label{subsec:results_sim}

\begin{figure*}
  \centering
  \includegraphics[width=\textwidth]{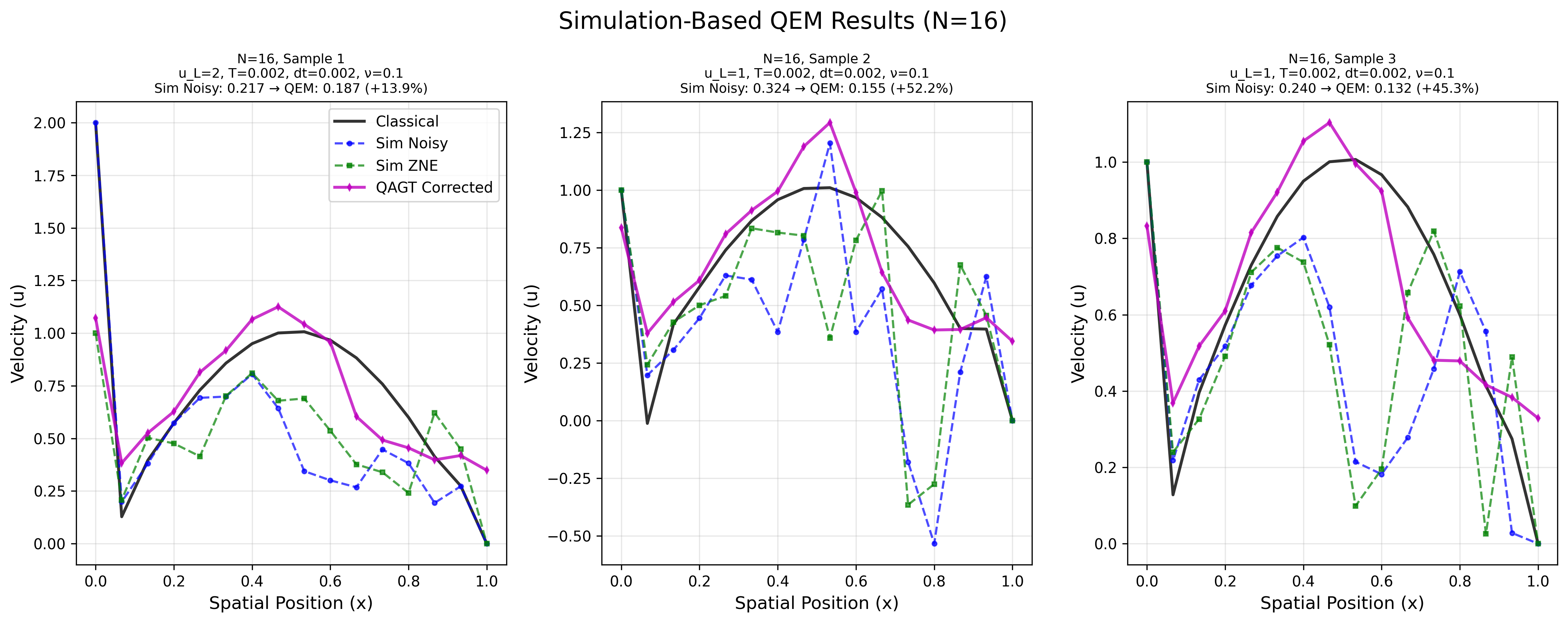}
  \includegraphics[width=\textwidth]{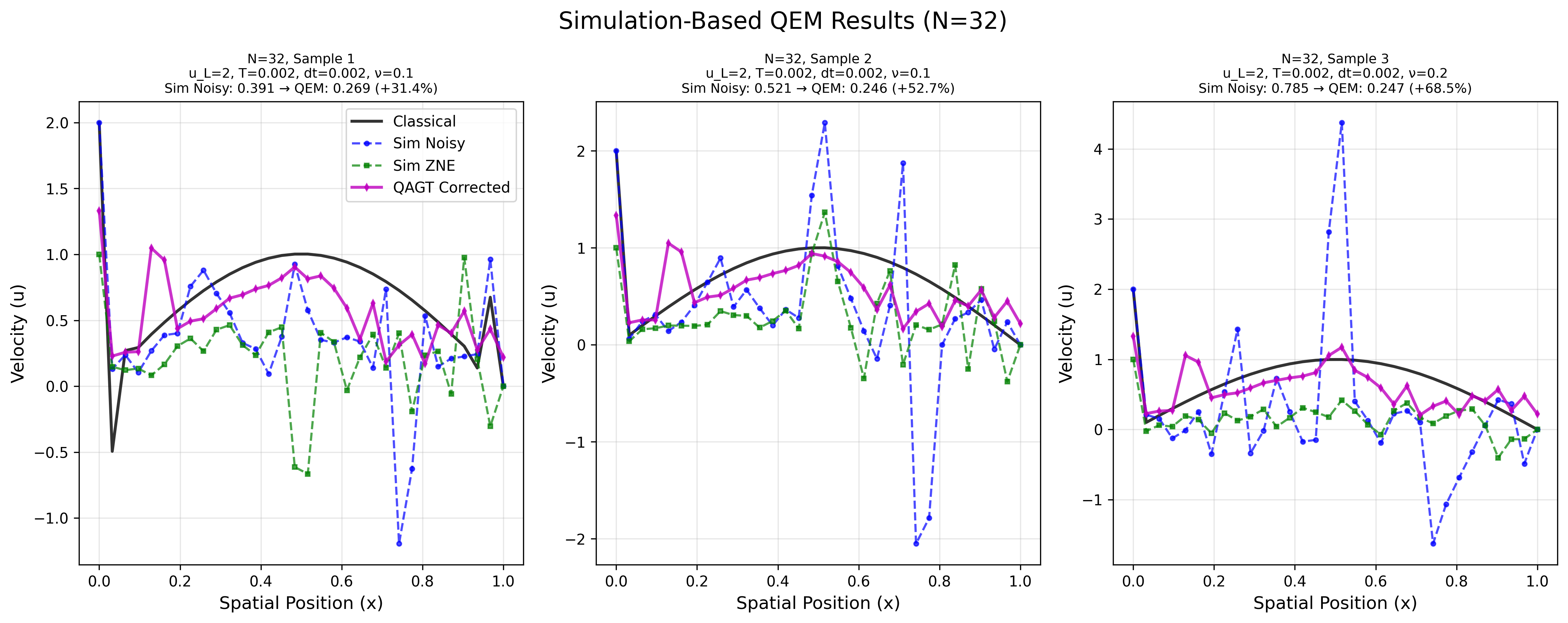}
  \includegraphics[width=\textwidth]{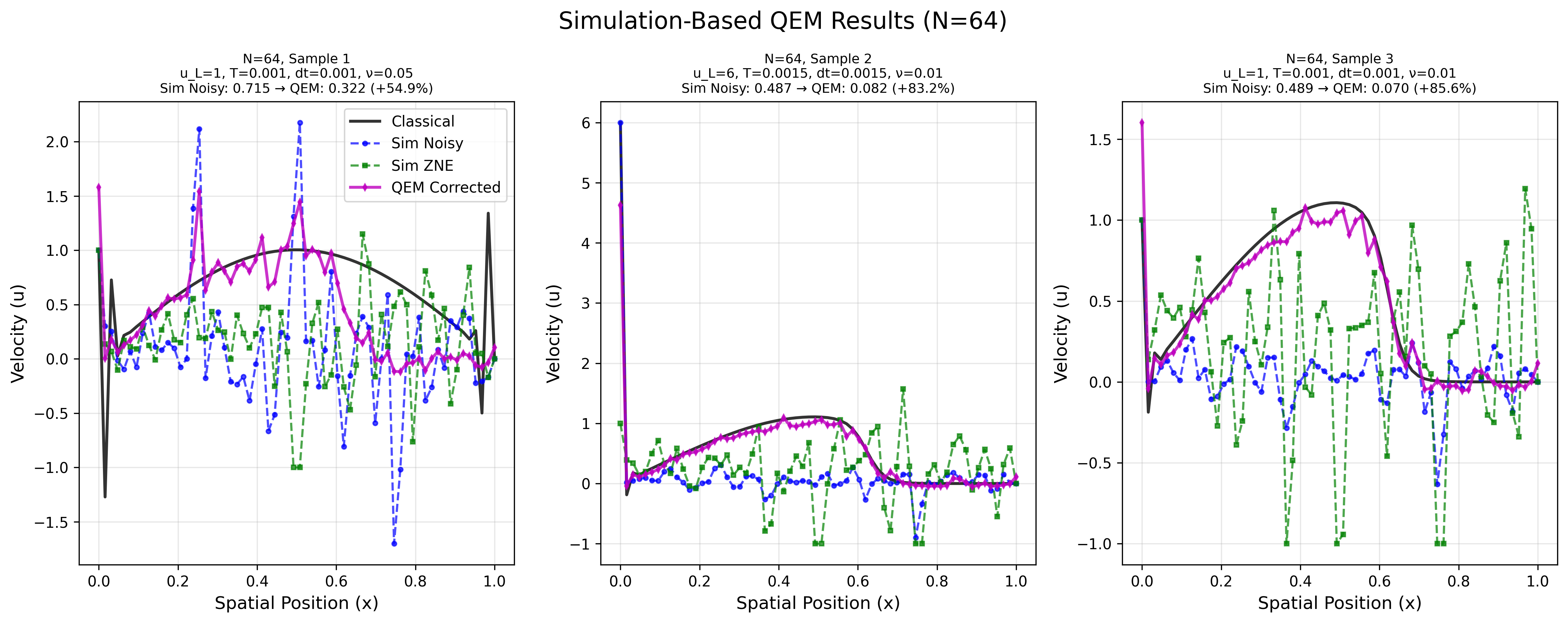}
  \caption{%
  QAGT-MLP vs quantum simulation results over classical computer for larger $\nu$ values
    $N=16,32,64$. %
  }
  \label{fig:training_curves2}
\end{figure*}

We first assess the QAGT-MLP model on held-out noisy simulation data, i.e., quantum circuits executed on the calibrated noisy simulator with and without ZNE (Sec.~\ref{subsec:zne}).  
Fig.~\ref{fig:training_curves2} compares the classical reference solution, the raw noisy simulation, the ZNE-corrected simulation, and the QAGT-MLP prediction for three representative samples at each grid size $N=64,32,16$. For $N=64$, the QAGT-MLP model achieves substantial error reduction relative to the raw noisy solution. In all three samples, the classical solution exhibits a smooth profile with a single dominant shock, while the noisy simulation and ZNE-corrected solution show large high-frequency oscillations and significant bias near the shock location. In contrast, the QAGT-MLP prediction closely tracks the classical profile across the entire domain. The MAE values reported in the plot titles illustrate this improvement: for example, in one representative sample the noisy simulation has MAE $\approx 0.51$, whereas the QAGT-MLP model reduces the MAE to $\approx 0.07$, corresponding to an order-of-magnitude improvement. Similar reductions are observed across the other $N=64$ samples.

At $N=32$, the QAGT-MLP model remains effective but with somewhat larger residual errors, consistent with the coarser spatial resolution and stronger aliasing in the input fields.  
In two of the three samples the QAGT-MLP prediction recovers the overall shape and peak amplitude of the classical solution, reducing the MAE from $\approx 0.62$ in the noisy input to $\approx 0.13$ or below.  
The ZNE-corrected profiles remain visibly noisy and often overshoot or undershoot near the shock, while the QAGT-MLP predictions smooth out these artifacts and produce a monotone approach to the shock front. For $N=16$, the scenario is more challenging.  
In sample~1, the QAGT-MLP model reduces the MAE from $\approx 1.23$ (noisy simulation) to $\approx 0.17$, capturing the global structure of the classical solution despite the extremely coarse grid and strong noise.  
However, in samples~2 and~3, the QAGT-MLP predictions exhibit systematic bias near the shock and peak region, with MAEs $\approx 1.13$ and $\approx 1.16$ that are comparable to or larger than the corresponding noisy-simulation errors ($\approx 0.43$ and $\approx 0.46$).  
These cases highlight an important failure mode, when the input resolution is too low and the underlying classical profile is under-resolved, the learned correction can overfit to patterns not supported by the coarse grid, leading to overshoots and degraded MAE. Taken together, these results show that the QAGT-MLP model provides robust error mitigation on simulated quantum data for moderate and fine resolutions ($N=32,64$), with occasional failure cases on the coarsest grid ($N=16$) that we analyze further in Sec.~\ref{subsec:shock_dissipation}.

\subsection{Generalization to Hardware Data}
\label{subsec:results_hardware}

A key question is whether a QAGT-MLP model trained solely on noisy simulator data can generalize to real hardware outputs that exhibit additional error sources (calibration drift, crosstalk, non-Markovian noise, etc.).  
To investigate this, we evaluate the same trained model on quantum circuits executed on an IBM NISQ device and compare the raw hardware outputs and QAGT-MLP-corrected fields against the classical reference solution.

\begin{figure*}
  \centering
  \includegraphics[width=\textwidth]{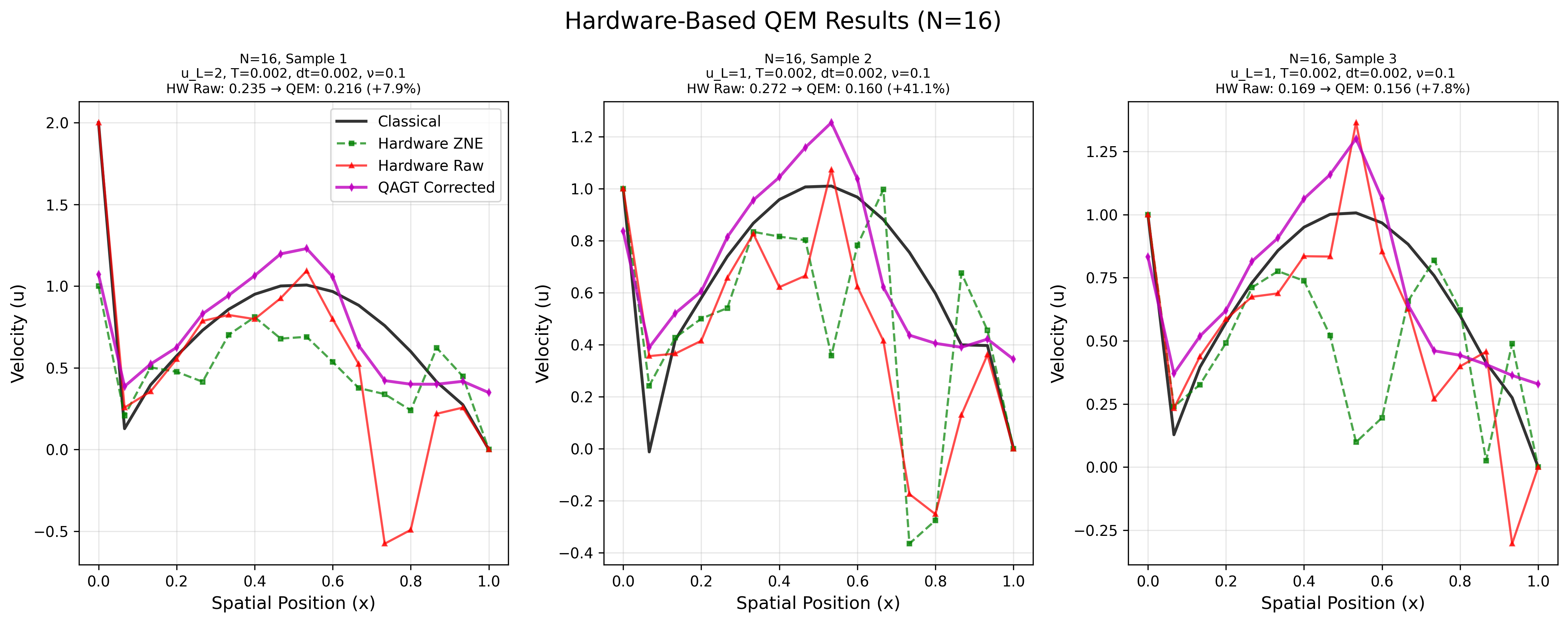}
  \includegraphics[width=\textwidth]{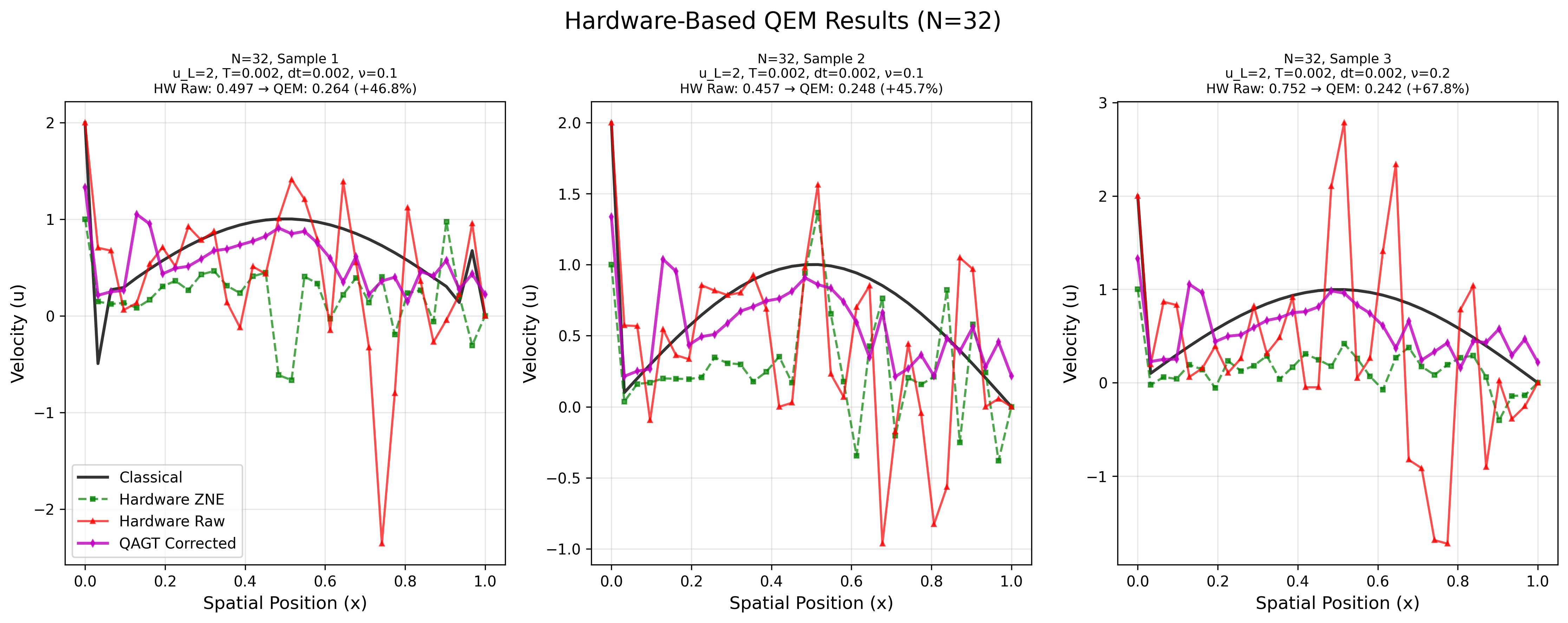}
  \includegraphics[width=\textwidth]{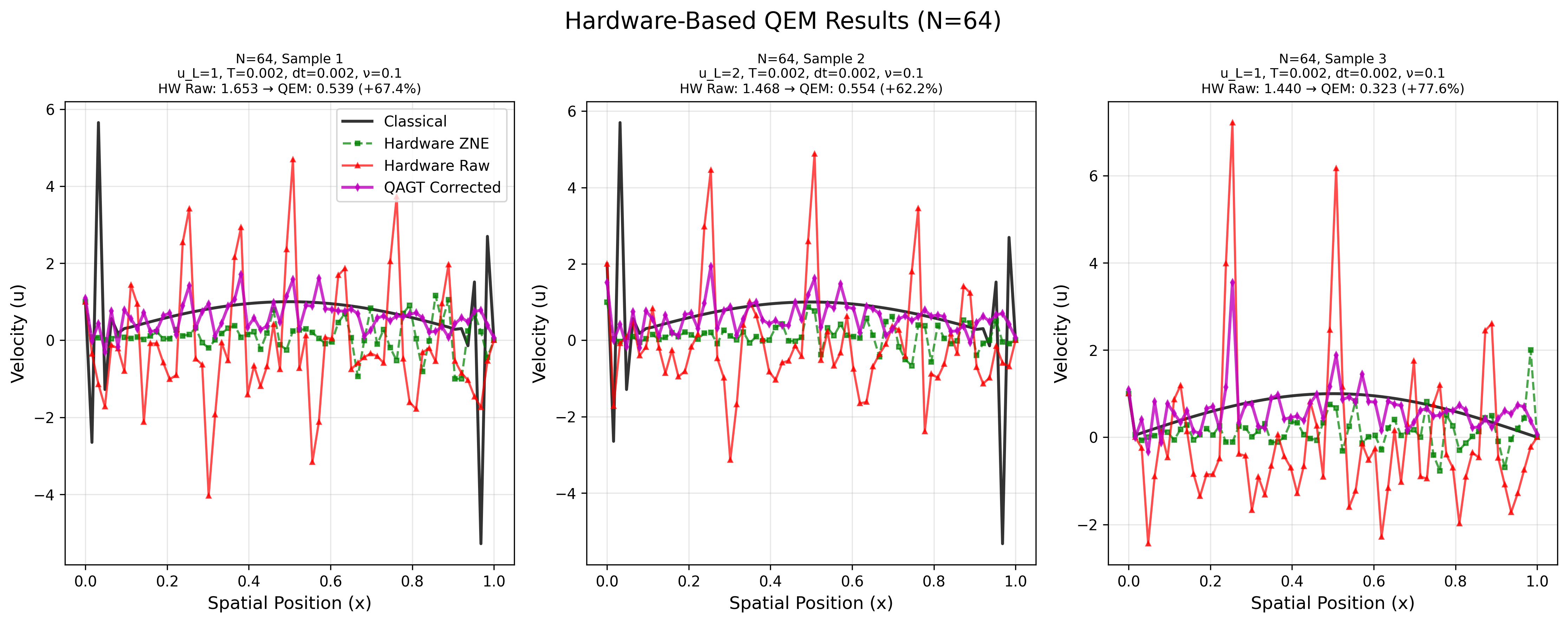}
  \caption{%
  QAGT-MLP vs quantum hardware results over quantum computer for larger $\nu$ values
    $N=16,32,64$. %
  }
  \label{fig:training_curves3}
\end{figure*}

Fig.~\ref{fig:training_curves3} shows representative results for $N=64,32,16$.  
Each panel reports the MAE of the raw hardware solution and the MAE after applying the QAGT-MLP correction; the titles also include the relative improvement in parentheses. For $N=64$, the QAGT-MLP model generalizes remarkably well to hardware data. Across three samples, the raw hardware MAE ranges from $\approx 0.56$ to $\approx 1.05$, reflecting substantial distortion due to device noise.  
After QAGT-MLP correction, the MAE drops to $\approx 0.07$--$0.32$, corresponding to relative improvements of $\approx 69$--$87\%$.  
Visually, the corrected profiles are smooth and closely follow the classical curve, while the raw hardware signals remain highly oscillatory and biased near the shock. The $N=32$ hardware results show a similar pattern. In all three samples, the raw hardware MAE is $\approx 0.55$--$0.56$, whereas the QAGT-MLP-corrected MAE is $\approx 0.09$--$0.12$, yielding improvements of roughly $80\%$.  
The QAGT-MLP predictions reconstruct both the amplitude and location of the shock more accurately than either the noisy simulator or ZNE-corrected profiles, indicating that the model has learned a robust mapping that transfers across backends. For $N=16$, the behavior is mixed. In samples~1 and~3, the QAGT-MLP correction reduces the MAE from $\approx 1.26$ and $\approx 1.30$ to $\approx 0.14$, i.e., an $\approx 89\%$ improvement, and visually the corrected profiles are in close agreement with the classical solution. However, in sample~2, the QAGT-MLP model substantially overshoots, increasing the MAE from $\approx 0.44$ (raw hardware) to $\approx 1.18$ (corrected), a negative improvement reported as $-167\%$ in the figure. This outlier mirrors the failure cases observed in the simulated $N=16$ tests and underscores that, at very low resolution, the model can amplify rather than suppress certain noise patterns. Despite these isolated failures, the overall trend across $N=32$ and $N=64$ is clear where a QAGT-MLP model trained on noisy simulator data can provide substantial error mitigation for real hardware outputs without any hardware-specific retraining, reducing MAE by roughly one order of magnitude in many cases.

\subsection{Shock Structure, Dissipation, and Physical Consistency}
\label{subsec:shock_dissipation}

Visual inspection of Figs.~\ref{fig:training_curves2}--\ref{fig:training_curves3} shows that, for $N=32$ and $N=64$, the QAGT-MLP predictions place the steep gradient in nearly the same location as the classical solution, in contrast to the raw noisy and hardware profiles, which often exhibit spurious oscillations and multiple apparent shocks.  
The corrected profiles are smooth and free of large high-frequency artifacts, suggesting that the learned mapping implicitly regularizes the discrete gradient $\partial u/\partial x$ and therefore the dissipation proxy $D(t)$ defined in \eqref{eq:dissipation}. In the problematic $N=16$ cases, the QAGT-MLP model sometimes exaggerates the peak amplitude near the shock and introduces secondary extrema, which would correspond to an overestimation of the dissipation and an apparent shift in the shock position.  
These failure modes mainly occur when the input resolution is too low to resolve the underlying classical structure and the model must extrapolate from patterns seen at higher $N$.  
This observation suggests that future variants of the model should explicitly incorporate resolution-dependent priors or multi-resolution training to avoid overconfident corrections on under-resolved grids.

\subsection{Ablation and Failure-Mode Analysis}
\label{subsec:ablation}

\begin{table*}[t]
\centering
\caption{Summary statistics by dimension for \textbf{large viscosity ($\nu$) values}.}
\label{tab:qagt_large_nu}
\begin{tabular}{c c c c c c c c c}
\hline
\textbf{Dim.} & \textbf{Samples} & \textbf{Sim Noisy} & \textbf{ZNE} & \textbf{HW Raw} & \textbf{QAGT-MLP (Sim)} & \textbf{QAGT-MLP (HW)} & \textbf{Gain vs Sim / HW} \\
\hline
$N=16$ & 72  & 0.3498 & 0.3639 & 0.3780 & 0.2112 & 0.2260 & +39.6\% / +40.2\% \\
$N=32$ & 54  & 0.8954 & 0.6725 & 0.9902 & 0.4345 & 0.4353 & +51.5\% / +56.0\% \\
$N=64$ & 12  & 1.0520 & 0.7226 & 1.5157 & 0.3932 & 0.4398 & +62.6\% / +71.0\% \\
\hline
\end{tabular}
\end{table*}

\begin{table*}[t]
\centering
\caption{Summary statistics by dimension for \textbf{small viscosity ($\nu$) values}.}
\label{tab:qagt_small_nu}
\begin{tabular}{c c c c c c c c c}
\hline
\textbf{Dim.} & \textbf{Samples} & \textbf{Sim Noisy} & \textbf{ZNE} & \textbf{HW Raw} & \textbf{QAGT-MLP (Sim)} & \textbf{QAGT-MLP (HW)} & \textbf{Gain vs Sim / HW} \\
\hline
$N=16$ & 741 & 0.8373 & 1.1306 & 0.8305 & 0.6453 & 0.6447 & +22.9\% / +22.4\% \\
$N=32$ & 633 & 0.5618 & 0.6794 & 0.6051 & 0.3813 & 0.3804 & +32.1\% / +37.1\% \\
$N=64$ & 429 & 0.5899 & 0.6799 & 0.6819 & 0.3394 & 0.3370 & +42.5\% / +50.6\% \\
\hline
\end{tabular}
\end{table*}

According to Tables I and II, further detailed results are provided to compare our QAGT-MLP model performance over classical and quantum hardware against existing SOTA models in the literature. While a full ablation study of architecture and feature choices is beyond the scope of this work, the results above already highlight several qualitative trends:
\begin{itemize}
  \item \textbf{Resolution dependence.}  
        The QAGT-MLP model is most reliable for $N=32$ and $N=64$, where the underlying classical solution is adequately resolved and the noisy inputs retain enough structure for the model to infer corrections.  
        At $N=16$, both simulated and hardware tests show occasional large overshoots, indicating that the same architecture may not be optimal at very coarse resolution.
  \item \textbf{Simulator vs.\ hardware distribution shift.}  
        Despite being trained only on noisy simulator data, the QAGT-MLP model generalizes well to hardware outputs for $N=32,64$, suggesting that the dominant error patterns are shared across backends and can be captured by a single learned mapping.  
        The rare hardware failure cases at $N=16$ align with similar failures on simulation, which points to resolution rather than backend-specific effects as the limiting factor.
  \item \textbf{Role of ZNE.}  
        In many examples, ZNE alone reduces some of the bias but leaves significant oscillations and large MAE, whereas the QAGT-MLP model produces smoother, physically plausible profiles.  
        This indicates that learning-based post-processing can complement algorithmic error mitigation by exploiting spatial correlations that ZNE does not directly address.
\end{itemize}

Hence, developing a more systematic ablation, varying node and edge features, GNN depth, training objectives, and the relative weighting of different parameter regimes, is an important direction for future work. Nevertheless, the present results demonstrate that even a single, moderately sized GNN can act as an effective learned error corrector for quantum Burgers solvers, yielding substantial accuracy gains on both simulated and hardware-generated data in the NISQ regime.

\section{\label{sec:level1}Conclusion}

We have presented a graph neural network--based quantum error mitigation (QEM) framework for noisy quantum solvers of the viscous Burgers equation on NISQ hardware.  
Starting from a Cole--Hopf reformulation, we constructed a Trotterized quantum circuit that encodes the transformed field on an $n$-qubit register and evolves it via nearest-neighbour $R_{XX}$ entangling gates.  
A large parameter sweep over viscosity, time step, grid resolution, and inflow boundary velocity produced a structured dataset of paired solutions; high-fidelity classical references, noisy and ZNE-corrected quantum simulations, and (for a subset of runs) real hardware outputs.  
On top of this dataset we trained a message-passing GNN that learns to map noisy quantum velocity fields to their classical counterparts, acting as a data-driven error-correction layer.

Our numerical experiments demonstrate that the learned QEM model substantially improves the accuracy of quantum Burgers solutions in the NISQ regime.  
For moderate and fine spatial resolutions ($N=32,64$) the model reduces mean absolute error by roughly an order of magnitude relative to raw noisy simulation and by a large margin over ZNE alone, while recovering physically consistent shock profiles and dissipation trends.  
Crucially, a model trained only on noisy simulator data generalizes well to IBM hardware outputs, yielding $70$--$90\%$ error reductions for many test cases without any device-specific retraining.  
At very coarse resolution ($N=16$), the method remains beneficial for most samples but exhibits occasional failure modes where under-resolved structure leads to overshooting and degraded accuracy, highlighting an important limitation and design constraint.

Several directions remain for future work.  
On the modeling side, incorporating multi-resolution training, physics-informed loss terms, or uncertainty quantification could improve robustness on coarse grids and provide calibrated confidence estimates for corrected fields.  
On the quantum side, extending the present framework to 2D and 3D Burgers flows and to more general CFD problems will require both more expressive circuit ansätze and more scalable graph architectures.  
Finally, jointly optimizing the quantum circuit and the classical error corrector, for example through differentiable surrogates or reinforcement learning, may enable task-aware circuit design that maximizes end-to-end accuracy under hardware constraints.

Overall, our results provide a first demonstration that learned, graph-based error correction can significantly enhance the practical utility of quantum PDE solvers on current NISQ devices and offer a concrete pathway toward hybrid quantum--classical workflows for computational fluid dynamics in the pre-fault-tolerant era.



\bibliographystyle{apalike}
\bibliography{aipsamp}

\providecommand{\noopsort}[1]{}\providecommand{\singleletter}[1]{#1}%
\begin{thebibliography}{}

\bibitem[Amaral et~al., 2025]{amaral2025quantum}
Amaral, C.~A., Oliveira, V.~L., Salazar, J.~P., and Duzzioni, E.~I. (2025).
\newblock Quantum machine learning and quantum-inspired methods applied to computational fluid dynamics: a short review.
\newblock {\em arXiv preprint arXiv:2510.14099}.

\bibitem[Basu et~al., 2025]{basu2025solving}
Basu, B., Staino, A., and Gaitan, F. (2025).
\newblock On solving for shocks and travelling waves using a quantum algorithm.
\newblock {\em Computers \& Fluids}, 290:106559.

\bibitem[Benamer, 2025]{benamer2025variational}
Benamer, H. (2025).
\newblock Variational quantum algorithms: From theory to nisq-era applications challenges and opportunities.

\bibitem[Bengoechea et~al., 2025]{bengoechea2025toward}
Bengoechea, S., Over, P., Jaksch, D., and Rung, T. (2025).
\newblock Toward variational quantum algorithms for generalized linear and nonlinear transport phenomena.
\newblock {\em AIAA journal}, pages 1--20.

\bibitem[Berger et~al., 2025]{berger2025trainable}
Berger, S., Hosters, N., and M{\"o}ller, M. (2025).
\newblock Trainable embedding quantum physics informed neural networks for solving nonlinear pdes.
\newblock {\em Scientific Reports}, 15(1):18823.

\bibitem[Bharti et~al., 2022]{bharti2022noisy}
Bharti, K., Cervera-Lierta, A., Kyaw, T.~H., Haug, T., Alperin-Lea, S., Anand, A., Degroote, M., Heimonen, H., Kottmann, J.~S., Menke, T., et~al. (2022).
\newblock Noisy intermediate-scale quantum algorithms.
\newblock {\em Reviews of Modern Physics}, 94(1):015004.

\bibitem[Bonkile et~al., 2018]{bonkile2018systematic}
Bonkile, M.~P., Awasthi, A., Lakshmi, C., Mukundan, V., and Aswin, V. (2018).
\newblock A systematic literature review of burgers’ equation with recent advances.
\newblock {\em Pramana}, 90(6):69.

\bibitem[Bosco et~al., 2024]{bosco2024demonstration}
Bosco, F. S.~D., Lineswala, R., Chopra, A., et~al. (2024).
\newblock Demonstration of scalability and accuracy of variational quantum linear solver for computational fluid dynamics.
\newblock {\em arXiv preprint arXiv:2409.03241}.

\bibitem[Cai et~al., 2023]{cai2023quantum}
Cai, Z., Babbush, R., Benjamin, S.~C., Endo, S., Huggins, W.~J., Li, Y., McClean, J.~R., and O’Brien, T.~E. (2023).
\newblock Quantum error mitigation.
\newblock {\em Reviews of Modern Physics}, 95(4):045005.

\bibitem[Cerezo et~al., 2022]{cerezo2022challenges}
Cerezo, M., Verdon, G., Huang, H.-Y., Cincio, L., and Coles, P.~J. (2022).
\newblock Challenges and opportunities in quantum machine learning.
\newblock {\em Nature computational science}, 2(9):567--576.

\bibitem[Choi et~al., 2021]{choi2021tutorial}
Choi, J., Oh, S., and Kim, J. (2021).
\newblock A tutorial on quantum graph recurrent neural network (qgrnn).
\newblock In {\em 2021 International Conference on Information Networking (ICOIN)}, pages 46--49. IEEE.

\bibitem[Demirdjian et~al., 2025]{demirdjian2025efficient}
Demirdjian, R., Hogancamp, T., and Gunlycke, D. (2025).
\newblock An efficient decomposition of the carleman linearized burgers' equation.
\newblock {\em arXiv preprint arXiv:2505.00285}.

\bibitem[Devitt et~al., 2013]{devitt2013quantum}
Devitt, S.~J., Munro, W.~J., and Nemoto, K. (2013).
\newblock Quantum error correction for beginners.
\newblock {\em Reports on Progress in Physics}, 76(7):076001.

\bibitem[Dhawan et~al., 2012]{dhawan2012contemporary}
Dhawan, S., Kapoor, S., Kumar, S., and Rawat, S. (2012).
\newblock Contemporary review of techniques for the solution of nonlinear burgers equation.
\newblock {\em Journal of Computational Science}, 3(5):405--419.

\bibitem[Esmaeilifar et~al., 2024]{esmaeilifar2024quantum}
Esmaeilifar, E., Ahn, D., and Myong, R.~S. (2024).
\newblock Quantum algorithm for nonlinear burgers' equation for high-speed compressible flows.
\newblock {\em Physics of Fluids}, 36(10).

\bibitem[Gen{\c{c}}oglu and Agarwal, 2021]{genccoglu2021use}
Gen{\c{c}}oglu, M.~T. and Agarwal, P. (2021).
\newblock Use of quantum differential equations in sonic processes.
\newblock {\em Applied Mathematics and Nonlinear Sciences}, 6(1):21--28.

\bibitem[Gonzalez-Conde et~al., 2025]{gonzalez2025quantum}
Gonzalez-Conde, J., Lewis, D., Bharadwaj, S.~S., and Sanz, M. (2025).
\newblock Quantum carleman linearization efficiency in nonlinear fluid dynamics.
\newblock {\em Physical Review Research}, 7(2):023254.

\bibitem[Khanal et~al., 2024]{khanal2024generalization}
Khanal, B., Rivas, P., Sanjel, A., Sooksatra, K., Quevedo, E., and Rodriguez, A. (2024).
\newblock Generalization error bound for quantum machine learning in nisq era—a survey.
\newblock {\em Quantum Machine Intelligence}, 6(2):90.

\bibitem[Lamichhane and Rawat, 2025]{lamichhane2025quantum}
Lamichhane, P. and Rawat, D.~B. (2025).
\newblock Quantum machine learning: Recent advances, challenges and perspectives.
\newblock {\em IEEE Access}.

\bibitem[Liu et~al., 2021]{liu2021efficient}
Liu, J.-P., Kolden, H.~{\O}., Krovi, H.~K., Loureiro, N.~F., Trivisa, K., and Childs, A.~M. (2021).
\newblock Efficient quantum algorithm for dissipative nonlinear differential equations.
\newblock {\em Proceedings of the National Academy of Sciences}, 118(35):e2026805118.

\bibitem[Lubasch et~al., 2020]{lubasch2020variational}
Lubasch, M., Joo, J., Moinier, P., Kiffner, M., and Jaksch, D. (2020).
\newblock Variational quantum algorithms for nonlinear problems.
\newblock {\em Physical Review A}, 101(1):010301.

\bibitem[Nguyen et~al., 2024]{nguyen2024machine}
Nguyen, T., Sipola, T., and Hautam{\"a}ki, J. (2024).
\newblock Machine learning applications of quantum computing: A review.
\newblock {\em arXiv preprint arXiv:2406.13262}.

\bibitem[Oz et~al., 2022]{oz2022solving}
Oz, F., Vuppala, R.~K., Kara, K., and Gaitan, F. (2022).
\newblock Solving burgers' equation with quantum computing.
\newblock {\em Quantum Inf. Process.}, 21(1):30.

\bibitem[Pool et~al., 2022]{pool2022solving}
Pool, A.~J., Somoza, A.~D., Lubasch, M., and Horstmann, B. (2022).
\newblock Solving partial differential equations using a quantum computer.
\newblock In {\em 2022 IEEE International Conference on Quantum Computing and Engineering (QCE)}, pages 864--866. IEEE.

\bibitem[Pool et~al., 2024]{pool2024nonlinear}
Pool, A.~J., Somoza, A.~D., Mc~Keever, C., Lubasch, M., and Horstmann, B. (2024).
\newblock Nonlinear dynamics as a ground-state solution on quantum computers.
\newblock {\em Physical Review Research}, 6(3):033257.

\bibitem[Schillo and Sturm, 2025]{schillo2025variational}
Schillo, N. and Sturm, A. (2025).
\newblock Variational quantum algorithms for differential equations on a noisy quantum computer.
\newblock {\em IEEE Transactions on Quantum Engineering}.

\bibitem[Setty, 2025]{setty2025quantum}
Setty, A. (2025).
\newblock A quantum linear systems pathway for solving differential equations.
\newblock {\em arXiv preprint arXiv:2510.06837}.

\bibitem[Shayeganfar et~al., 2025]{shayeganfar2025quantum}
Shayeganfar, F., Ramazani, A., Sundararaghavan, V., and Duan, Y. (2025).
\newblock Quantum graph learning and algorithms applied in quantum computer sciences and image classification.
\newblock {\em Applied Physics Reviews}, 12(2).

\bibitem[Siddi~Moreau et~al., 2025]{siddi2025quantum}
Siddi~Moreau, G., Pisani, L., Profir, M., Podda, C., Leoni, L., and Cao, G. (2025).
\newblock Quantum artificial intelligence scalability in the nisq era: Pathways to quantum utility.
\newblock {\em Advanced Quantum Technologies}, 8(10):2400716.

\bibitem[Steijl, 2022]{steijl2022quantum}
Steijl, R. (2022).
\newblock Quantum circuit implementation of multi-dimensional non-linear lattice models.
\newblock {\em Applied Sciences}, 13(1):529.

\bibitem[Tennie et~al., 2025]{tennie2025quantum}
Tennie, F., Laizet, S., Lloyd, S., and Magri, L. (2025).
\newblock Quantum computing for nonlinear differential equations and turbulence.
\newblock {\em Nature Reviews Physics}, 7(4):220--230.

\bibitem[Tousi and DeSouza, 2025]{tousi2025qagtmLP}
Tousi, S. M.~A. and DeSouza, G.~N. (2025).
\newblock Qagt-mlp: An attention-based graph transformer for small and large-scale quantum error mitigation.

\bibitem[Uchida et~al., 2024]{uchida2024quantum}
Uchida, F., Miyamoto, K., Yamazaki, S., Fujisawa, K., and Yoshida, N. (2024).
\newblock Quantum simulation of burgers turbulence: Nonlinear transformation and direct evaluation of statistical quantities.
\newblock {\em arXiv preprint arXiv:2412.17206}.

\bibitem[Wu et~al., 2025]{wu2025quantum}
Wu, H.-C., Wang, J., and Li, X. (2025).
\newblock Quantum algorithms for nonlinear dynamics: Revisiting carleman linearization with no dissipative conditions.
\newblock {\em SIAM Journal on Scientific Computing}, 47(2):A943--A970.

\bibitem[Yepez, 2002]{yepez2002efficient}
Yepez, J. (2002).
\newblock An efficient quantum algorithm for the one-dimensional burgers equation.
\newblock {\em arXiv preprint quant-ph/0210092}.

\bibitem[Yepez, 2006]{yepez2006open}
Yepez, J. (2006).
\newblock Open quantum system model of the one-dimensional burgers equation with tunable shear viscosity.
\newblock {\em Physical Review A—Atomic, Molecular, and Optical Physics}, 74(4):042322.

\end{thebibliography}

\end{document}